\documentclass{article}
\usepackage{amsmath}
\usepackage[utf8]{inputenc}
\usepackage{amsfonts,url,epsfig,breakurl}
\usepackage{geometry}
\usepackage{sectsty}
\usepackage[normalem]{ulem}
\usepackage{dirtytalk}
\usepackage{float}

\usepackage{multirow}
\usepackage{array}
\newcolumntype{P}[1]{>{\centering\arraybackslash}p{#1}}

\newcolumntype{M}[1]{>{\centering\arraybackslash}m{#1}}

\usepackage[style=apa,sortcites=true,sorting=nyt,backend=biber]{biblatex}
\DeclareLanguageMapping{american}{american-apa}

\newcommand\blfootnote[1]{%
  \begingroup
  \renewcommand\thefootnote{}\footnote{#1}%
  \addtocounter{footnote}{-1}%
  \endgroup
}

\title{\bf He, she, they: Using sex and gender in survey adjustment\vspace{.1in}}
\author{Lauren Kennedy\footnote{Department of Econometrics and Business Statistics, Monash University} \and Katharine Khanna\footnote{Department of Sociology, Columbia University} \and Daniel Simpson\footnote{Department of Statistical Sciences, University of Toronto} \and Andrew Gelman\footnote{Department of Statistics and Department of Political Science, Columbia University}\and Yajun Jia\footnote{Center on Poverty and Social Policy, Columbia University}\and Julien Teitler\footnote{School of Social Work, Columbia University}\vspace{.1in}}

\date{23 Mar 2022}

\begin{document}\sloppy
\maketitle

\begin{abstract}

Accounting for sex and gender is a challenge in social science research. While other methodology papers consider issues surrounding appropriate measurement, we consider the problem of adjustment for survey nonresponse and generalization from samples to populations in the context of the recent push toward measuring sex or gender as a non-binary construct.  This is challenging not only in that response categories differ between sex and gender measurement, but also in that both these attributes are potentially multidimensional. We reflect on similarities to measuring race/ethnicity before considering the ethical and statistical implications of the options available to us. We present a simulation study to understand the statistical implications under a variety of scenarios, and demonstrate the application of the decision process with the New York City Poverty Tracker. Overall, we conclude not with a single best recommendation for all surveys but rather with an awareness of the complexity of the problem and the benefits and weaknesses of different approaches. 

\end{abstract}
\blfootnote{We thank the U.S. Office of Naval Research, National Science Foundation, National Institutes of Health, the Natural Sciences and Engineering Research Council of Canada, the Canadian Research Chair program and the Institute of Education Sciences for their partial support of this work.
KK would like to acknowledge support from the National Science Foundation Graduate Research Fellowship, grant number: DGE-1644869.\\ \*Corresponding author. Email: {\rm Lauren.Kennedy1@monash.com}}

\section{Introduction} 

The challenges faced by surveys are increasing. \textcite{pew_responserates} report that the response rate of their surveys has continued to drop, a pattern that many surveys face. This creates a potential threat for representation, as a lower response rate can increases the impact of heterogeneous response between demographics. A lower response rate also increases costs, leading to a surge in the use of non-probability samples and panel-based sampling. At the same time, surveys are often asked to do more with less data, generally in the form of creating estimates of smaller demographics and geographic areas. 

Usual weighting or poststratification-based adjustments require that demographic counts of adjustment variables must be known in the population, and this is a problem when demographic questions are asked with different wording or different measurement categories compared to the census. One solution to this is to mirror the census demographic questionnaire to ensure comparable measurements. However, this is not always desirable or possible because rather than the survey question choice being driven by the needs of the survey, instead it is driven by available population information.

One instance of this is the measurement and use of sex or gender in surveys. The U.S. census measures sex, but increasingly surveys have moved to measurement of gender as a key demographic. This creates challenges in poststratification adjustment as two different but related demographics need to be harmonized. In addition, the measurements themselves may not be analogous because while sex is typically as a binary category, gender can be measured with more responses or even as a continuum. 
There may be no good way to resolve sex and gender measurement in surveys. However, it is an important problem both structurally and  to individuals. Despite the strong overlap between the two variables, there is no simple mapping from sex to gender that works for the entire population. This is true for binarized gender and sex categories but becomes especially urgent once we consider intersex, transgender, nonbinary, and other categories. This affects survey research, not just for surveys that directly concern gender and sex roles, but also in nonresponse adjustment which is common in many surveys.  For example, it has long been standard practice for political polls to adjust for sex, along with other variables such as age, ethnicity, and education \parencite{voss1995polls}, in part because women have traditionally responded to surveys at a higher rate than men.  

In this manuscript we first overview the broader context of sex and gender measurement in surveys, and draw links to similar topics on race/ethnicity measurement. We then tackle the potential solutions to the specific problem of sex and gender measurement using a statistical and ethical lens. Lastly we finish with an application focused on estimating poverty and other types of disadvantage in New York City. 

\subsection{Changes in definitions and measurements of sex and gender}

Since the 1950s, psychology has distinguished between sex and gender \parencite[see][for a historical overview]{muehlenhard2011distinguishing}, with gender becoming increasingly considered more relevant for social research \parencite{basow2010changes}. \textcite{glasser2008vague} also cite instances where gender and sex are \say{vague, conflated and apparently synonymous} (p.\ 345), particularly when they are measured in binary terms.

However, when measuring gender with simply two categories, there is a failure to capture the unique experiences of those who do not identify as either male or female, or for those whose gender does not align with their sex classification. To rectify this, it is recommended that researchers include a more diverse set of possible responses. \textcite{cameron2019gender} recommend open responses for gender, but for the particular problem of survey adjustment we focus on three response categories, as survey adjustment typically works with discrete variables.

The move to measuring gender with (at least) three categories has highlighted an already existing problem. To adjust by gender, we must first create some sort of mapping from gender to sex. When gender was measured in binary format, male could be mapped to male and female mapped to female (again highlighting the suggestion of \textcite{glasser2008vague} for synonymous use). The addition of a non-binary category, however, forces us to consider mapping sex to gender more generally.\footnote{We use non-binary as a category name for those who identify as non-binary, agender, gender fluid, and other gender identities outside of female and male. Although ``other'' is commonly used for this category, we specifically do not use this term to avoid othering those who do not identify as male or female.} Although some census bureaus are beginning to measure both gender identity and sex assigned at birth \parencite{cancensus2021}, this is a recent development limited to only some countries. It is likely that this challenge will remain relevant for some time. 

For survey weighting techniques where only gender is adjusted for, one temptation is to simply give the ``non-binary'' respondents an average weight. This avoids imputing sex or gender, but implies that the weight for non-binary respondents should be dependent on the relative ratio of the over/undersampling of male and female respondents, and there is no reason to believe this is the case. In addition, this continues to perpetuate the conflation of gender and sex. Aside from this, in many surveys as we adjust for more and more variables, we increasingly rely on methods like raking. For these methods, each category in the sample has to be matched to a category in the population. Similarly, for methods such as multilevel regression and poststratification \parencite{gelman1997poststratification, park2004bayesian}, one would either need to (stochastically) impute a binary variable in the sample or else construct a model on the expanded space with three or more options.

In this manuscript, our aims are to consider potential options available. Key concerns are {\em ethics} (respecting the perspectives and dignity of individual survey respondents), {\em accuracy} (for estimates of the general population and for subpopulations of interest), {\em practicality} (using more complicated procedures only when they serve some useful function), and {\em flexibility} (anticipating future needs).  In addition, the effort spent thinking through the coding of sex and gender can be useful when considering other survey responses that involve complex measurement (such as race/ethnicity and religion).

\subsection{Implications}

For online surveys, which rely solely on poststratification to adjust an unrepresentative sample to the population of interest, this decision will be particularly of importance. In \textcite{kennedy2020know}, we attempted to adjust an online survey to the U.S. population. The survey, which was developed by psychologists, measured gender with three categories, but the U.S. census measured sex as two. We removed those who responded other from the dataset for pedagogical reasons. However, this began our awareness of this problem and of the inappropriateness of the solution we had used. We noted this in the cited manuscript and began the more considered investigation presented here.

Of course it is not simply psychological research that has faced this challenge. For example, in 2020, the New York City Longitudinal Survey of Wellbeing, also known as the Poverty Tracker \parencite{PT2018, povertytracker} recognized that their existing measurement of gender as exclusively male or female didn't reflect their desire to respect respondents' identity. They moved to measuring gender with three categories (male, female, and an open-ended other option), which again raised the issue of how to appropriately weight the sample when only sex is known at a population level. 

Moreover, increasingly it is becoming apparent that there will be no one-size-fits-all measurement solution, which means there can also be no automatic statistical solution. For example, the U.K. office for national statistics \parencite{uk_census} is recommending the use of a second question that asks whether the respondent identifies as the same gender as sex registered at birth, and free response if not. This is similar to the differences in race/ethnicity questions between different countries and will make cross-national and cross-time research difficult. 

Indeed, the population as measured in the census may not even be the target population of interest. If the population of interest is a community of LGBTQIA+ individuals, then it is likely that there is a higher proportion of non-binary individuals when compared to the population of the U.S. In this case, appropriate poststratification of a sample to the population counts will potentially have a larger impact on overall and subgroup estimates. 

In considering this, it is worth spending a moment considering why we adjust for sex or gender in surveys at all. These reasons represent a myriad of sociological and statistical concerns. Sociological concerns range from historic under-representation, to current response patterns, to ensuring that those who are discriminated against and have lower power in society are represented. Statistical concerns reflect on the relationship between sex/gender and many outcomes of interest.

\subsection{Definitions of sex and gender}

Gender and sex have been defined in different ways at different times.  We use the definitions provided by the \textcite{SexGenderDef}:

\begin{quotation}\noindent
``Sex refers to a set of biological attributes in humans and animals. It is primarily associated with physical and physiological features including chromosomes, gene expression, hormone levels and function, and reproductive/sexual anatomy. \dots

Gender refers to the socially constructed roles, behaviours, expressions and identities of girls, women, boys, men, and gender diverse people. It influences how people perceive themselves and each other, how they act and interact, and the distribution of power and resources in society.\ \dots''
\end{quotation}

For most people, the question of which construct is being measured is moot---they would respond the same regardless. However, for the subset of people for whom sex and gender differ, we realize that both variables are multidimensional.  For example, in the above definition, sex has at least four dimensions (chromosomes, gene expression, hormones, and anatomy) as does gender (roles, behaviors, expressions, and identities). 

In this paper, we consider two challenging scenarios:
\begin{enumerate}
    \item A survey has three or more response categories to elicit gender, but we wish to poststratify to a population where sex is measured as either male or female.  This will arise, for example, when raking to the U.S. census.
    \item We want to combine data from multiple surveys that ask sex or gender in different ways, or allow different responses to these questions.
\end{enumerate}

These problems are not unique to sex and gender.  The first scenario, for example, arises for other survey weighting variables such as race or ethnicity, while the second scenario arises for variables such as income, which can be measured and constructed in different ways in different surveys.

Unless the survey or census question is very specific, responses can capture a mix of all the dimensions of sex and gender listed above.  For example, the 2020 U.S. census asks, ``What is Person 1's sex? Mark ONE box: male or female.''  The subset of people who might have difficultly responding to this question can choose what aspect of sex or gender they would like to use in their response.  Even though the variable is labeled as sex, the response can include some aspects of gender, as is there some freedom in what biological sex characteristics are used in the response.

Some large surveys are moving to measure both gender and sex assigned at birth. For instance, the Canadian census is planning to measure sex at birth and gender identity separately in the 2021 census \parencite{cancensus2021}. The General Social Survey also began this practice in 2018 \parencite{smith2019transgender,lagos2021}, although sex and gender are still confused, with responses to \say{What is your current gender?}\ referred to as \say{SEXNOW}. This does not resolve all challenges, though, even for surveys conducted in Canada, as sex at birth does not capture all the dimensions of biological sex, nor does the response to a gender identity question capture all dimensions of gender-related roles, behaviours, expressions, and identities, such that measurement differences can still occur. Yet by actively measuring both sex and gender constructs, the Canadian census makes adjustment considerably simpler. 

\subsection{Measuring identification in survey research: Studies of race/ethnicity and sex/gender}
Race and ethnicity, much like sex and gender, are socially constructed identities that are constituted through a range of attributes (skin colour, facial features, country of birth, racial self-identification, language, and culture). Survey research on race and ethnicity has similarly grappled with the challenges of measuring these identities in meaningful and consistent ways. As \textcite{roth2016multiple} notes, ``With the word `race’ used as a proxy for each of these dimensions, much of our scholarship and public discourse is actually comparing across several distinct, albeit correlated, variables'' (p.\ 1310). The multidimensionality of identities like race and gender suggests that precisely what we measure, and precisely how this measure is interpreted by the respondent,  can have profound impacts on our findings.

To account for this multidimensionality, some research has examined the advantages of employing multiple measures within a single survey to better understand the implication of each dimension for social inequality. For example, by using multiple measures of race within the same survey to disentangle what each dimension represents and how identification across the measures differs, \textcite{saperstein2016} have argued that ``the relative importance of various dimensions of race likely depends on the outcome in question.'' With these issues in mind, it remains an open question how researchers should proceed when they want to compare across data sets that inconsistently employ single measures of race (or gender). The importance of the outcome of interest in determining the best operationalization to measure inequality suggests that there may not be a one-size-fits-all solution for how to merge two such data sets.

Prior research has also shown that racial identity is not always stable over time. In a study comparing the 2000 and 2010 U.S. censuses, \textcite{liebler2017america} find that change in racial self-identification is common, especially among those who do not fall squarely within the single-race White, Black, or Asian categories. Moreover, racial self-identification does not always match others' perceptions of one's own race. This matters not only because others' perceptions may be important determinants of inequality, independent of racial self-identification, but also because racial contestation itself is an increasingly prevalent social process that contributes to strength of racial group commitment and identification \parencite{vargas2016documenting}, which studies have shown is associated with a range of social attitudes and behaviors \parencite{abascal2015us,ellemers1999social,ellemers2002self}.

Measurement of sex and gender has encountered similar problems, but fewer studies have examined the consequences of survey items that measure sex and gender in different ways. \textcite{bittner2017sex} note, ``The principal problem is the conflation of gender with sex in survey research. Consequently, gender is typically treated as a dichotomy, with no response options for androgynous gender identities, or indeed degrees of identification with masculine or feminine identities'' (p.\ 1019). Inconsistency in which attribute is measured (sex or gender) and the range of responses available poses challenges for researchers who want to combine or compare across multiple surveys. One reason for the limited study of sex and gender minorities is that they represent a smaller fraction of the general population, compared to many racial or ethnic minorities.  Properly adjusting for sex and gender classification becomes more relevant when studying targeted subgroups or when measuring low-frequency attitudes or behaviors in terms of statistical bias, but for the dignity of respondents it is always relevant. 

%\section{Weighting}
%Survey weights are traditionally constructed by multiplying weights from different poststratification or weighting steps.  For example, some political polls weight for sex $\times$ ethnicity and age $\times$ education, raking over each of these two-dimensional tables.  If sex has two categories and ethnicity has four categories, then there are eight cells in sex $\times$ ethnicity, and poststratification gives eight weights.  But what do we do with respondents who refuse to report sex or who give an alternative response?

%Firstly there is a difference between not responding to this question (missingness) and choosing a third option in response to this question. 

%If the refuse-to-reports are a small fraction of the population, then the following strategy should not cause a problem: first perform weighting using the people whose sex and ethnicities fall into the categories defined by the census, and then use average weights for people who did not respond.  For example, a person who responds White to the ethnicity question but does not respond to the sex question would be given a weight that is an average of the weights for White women and White men.

%This procedure should be fine for estimating population averages or totals, and it does not raise the ethical concerns associated with removing sex/gender nonrespondents from the sample, or with imputing them to male or female.

\subsection{Missingness versus non-binary genders}

Imputation of demographic variables is not unusual in surveys, and can be necessary to create survey weights. There is a difference between a respondent whose response to gender is missing, and one who actively chooses a category that is neither male or female.  We also need to account for transgender people who choose a male or female gender response that is the opposite of their recorded sex. For missing respondents, imputation is a procedure that assigns potential values to the respondent had they responded, generally using a model or some other information and in such a way to respect uncertainty of these potential values. If the missingness is truly missing at random, then this is not particularly unethical, but it should be remembered that respondents who do not identify as male or female may choose to skip this question in protest or because they are not sure how to respond. In this case, nonresponse is disproportionately akin to answering as neither male or female. For those respondents who are given the option of more than two categories, it is clear that they have actively indicated that they do not identify as either male or female, and so they shouldn't be identified as such. 

\section{Poststratification and gender measurement}

\begin{table}
\centering
\begin{tabular}{ l ccc }
  & Impute & Remove & Impute \\
  & sample  & respondents &  population \\
 \hline
Assume population distribution & Yes & No & Yes \\
Model population distribution using auxiliary data & Yes & No & Yes \\
Estimate gender using auxiliary information & Yes & No & No \\
Impute all non-male as female & Yes & No & No \\
Remove all non-binary respondents & No & Yes & No
\end{tabular}
\caption{\em Possible options for scenario 1. Columns represent potential facets of ethical consideration, while rows represent possible facets of statistical consideration. The cells represent whether it was possible to address these considerations together.}
\label{tab:pot_options_scenario1}
\end{table}
 
We consider a scenario that is increasingly common within the United States. A survey measures gender with three response categories (male, female, and non-binary), but the population data to which we would like to poststratify to measures sex with two response categories (male and female). In this scenario there are multiple issues at hand. Firstly, as we've discussed, sex and gender are separate and distinct constructs. Secondly, even if they were the same construct, they are measured with different potential categories. We create a matrix of potential solutions in Table~\ref{tab:pot_options_scenario1}. 

One of the challenges of considering the potential options is the interaction between statistical and ethical issues. Typically, scientists are trained in either one or the other, but rarely are we educated in detail on the intersection between the two. In this manuscript, we are interested in both, and so we discuss the potential options first in terms of their potential ethical considerations before considering the statistical considerations. 

\subsection{Ethical concerns}

There are ethical concerns with the collection and protection of gender and sex in surveys. These issues include data sensitivity and security \parencite{holzberg2017assessing} and the purpose of collecting such information \parencite{federal2016current}. Here we assume that collected gender is necessary for adjustment and to ensure adequate representation across genders, and that security risks can be mitigated appropriately. 
 
\subsubsection*{Imputing sample sex}
This method involves imputation using the gender reported by an individual in the sample to predict their potential response for their sex. In two of the three potential methods, this will involve directly imputing those who respond male as male, those who respond female as female, and those who respond non-binary as either male or female. The remaining method (using auxiliary data) does allow the potential to impute female sex as male gender and vice versa. 

As researchers we can be clear that we are imputing a potential answer to a binary sex question from a non-binary gender item. However, the current confusion between sex and gender within academic literature makes this practice appear as statistical misgendering, where an individual is incorrectly referred to as a gender with which they do not identify \parencite{misgenderdef}. The right to self-identify is protected by law in some jurisdictions \parencite{canada_humanrights}, with misgendering identified as a form of discrimination. In this scenario respondents have explicitly identified as neither male nor female, yet in our statistical analysis we are assigning them to this male/female dichotomy. In survey research as in other aspects of life, people have the right to define how they are identified.  This continues to be true when multiple imputation is used to represent the uncertainty in the classification because, as \textcite{Keyes2018Misgendering} states, \say{an error rate that disproportionately falls on one population is not just an error rate it is discrimination}. It is algorithmic injustice \parencite{noble2018algorithms}.

It could be argued that rather than imputing an individual's gender, we are instead imputing their expected response to a question as posed by the census (What is your sex?\ M/F). This may be the methodologist's intent, however, it is impossible to ensure that it is understood by users of the data, the survey respondents, and the populations affected by the survey analysis. In addition, this may be completed post collection without respondents' explicit consent, which creates further ethical concerns. 

Another challenge to this technique is the consideration of imputation error. Is there a difference to imputing potential sex responses based on demographic patterns when compared to other less formal imputation procedures, such as identification by interviewer or complex features of other covariates collected with machine learning methodologies? Imputation by demographic proportions reinforces the statistical need to know the proportions of different cells for survey adjustment, and seems analogous to using a method such as raking to impute potential cell proportions when only margins are known. Imputation by interviewer or through complex machine learning techniques has a greater emphasis on imputing the individual, which as we have already discussed is potentially unethical and  discriminatory. 

\subsubsection*{Remove respondents}

This method involves removing respondents who do not identify as either male or female from the sample when constructing survey weights. This technique is easily communicated to respondents, data users, and the wider public. It avoids the potential misgendering issues described in the previous section on imputing sex by avoiding assigning sex altogether. 

However, this method means that the responses of non-binary individuals are not counted for any analysis where the analyst wishes to make population generalizations. Participating in a study has, at a minimum, a time cost (and can potentially have others costs) that cannot be justified if non-binary respondents' data is not used. Moreover, this structural exclusion is a form of discrimination against non-binary individuals. If one purpose of surveys is to ensure equal and fair representation, then this method actively prevents non-binary respondents from having this opportunity. When it comes to population mean estimates, removing non-binary respondents is roughly equivalent to assuming that their responses would essentially be the weighted mean between male and female estimates.

\subsubsection*{Impute population values}

The ethical considerations associated with imputing the population might appear to mirror that of the sample, but there are additional nuances. To understand this, consider two different scenarios. 

The first scenario is a large population ($N \to \infty$) that has been summarised by a number of discrete categories such that the number of individuals who fall within each combination is labelled $N_j$, where $N_j$ is also sufficient large. We assume that each $N_j$ can be further split into $N_{j,{\rm sex \,=\, f}}$ and $N_{j,{ \rm sex \,=\, m}}$. In this scenario when we refer to ``imputing the population,'' we refer to using either using either a model or known distributions of response to split the cell $N_{j,{\rm sex \,=\, f}}$ into $N_{j,{\rm gender \,=\, f}}$, $N_{j,{\rm gender \,=\, non-binary}}$, $N_{j,{\rm gender \,=\, m}}$ and $N_{j,{\rm sex \,=\, m}}$ into $N_{j,{\rm gender \,=\, f}}$, $N_{j,{\rm gender \,=\, non-binary}}$, $N_{j,{\rm gender \,=\, m}}$. A simpler version would be to split the cell $N_{j,{\rm sex \,=\, f}}$ into $N_{j,{\rm gender \,=\, f}}$, $N_{j,{\rm gender \,=\, non-binary}}$ and $N_{j,{\rm sex \,=\, m}}$ into $N_{j,{\rm gender \,=\, non-binary}}$, $N_{j,{\rm gender \,=\, m}}$. For sufficiently large cells it is clear that this does not involve imputing any particular person's gender, which avoids the previous misgendering challenges.

The second scenario we consider is a relatively small population such that $N_j$ contains only a small number or even one individual. Unlike in the previous scenario, we can no longer ignore the finite sample effects of this imputation.  This raises multiple issues. The first is that it returns us to the original problem of imputing a specific person's gender rather than an abstract expectation for a cell. The second is that it becomes more difficult to split a particular cell. For instance, if in the population a cell contains only a single individual labelled as male sex, but we wish to impute their gender, it is difficult to reflect the uncertainty of their gender, due to the relative size of expectation for each potential gender option. A third issue is that the potential error rate is difficult to calculate with a small proportion of the population. 

\subsection{Statistical concerns}

Although we cannot consider statistical concerns without considering also ethical concerns, we use this section to describe potential statistical options. 

\subsubsection*{Assume known population proportions}

Assuming that we need to impute population data, perhaps the simplest approach is to use auxiliary information about the estimated table of gender distribution to impute gender at the population level. To do this we would assume a certain gender distribution in the population (without a census, we cannot know the proportions, but we set to 49\% female, 49\% male, and 2\% non-binary in this manuscript). We would then use this distribution to add gender to the poststratification table. It's likely that we would split the cell $N_{j,\,{\rm sex\, =\, f}}$ into $N_{j,\,{\rm gender\, = \, f}}$, $N_{j,\,{\rm gender \,=\, non-binary}}$ and $N_{j,\,{\rm sex\, =\, m}}$ into $N_{j,\,{\rm gender\, =\,  non-binary}}$, $N_{j,{\rm gender\, =\, m}}$. There will be some bias from respondents who identify as male sex and female gender and vice versa. This procedure also doesn't allow any uncertainty in the imputed gender counts to be included in the overall model or estimates. 

\subsubsection*{Use auxiliary data}

This method is similar to the previous method. It can be used in either the sample or the population. If used in the sample, a model predicting sex given other variables is created and for each participant their expected sex is imputed. If used in the population, a model predicting gender given other variables is created and each poststratification cell $N_j$ is imputed based on the expected proportion of male, female, and non-binary respondents.

One use case for auxiliary data is if a completely separate reference auxiliary data set measures both sex and gender, as well as a number of other demographics. This model is used to model either gender by sex and demographics (if imputing gender in the population) or sex by gender and demographics (if imputing sex in the sample). The benefit of this is that it simply allows for better imputation at the cell level to encompass demographic differences in gender identity. This is not imputing sex or gender in the auxiliary data, but rather modeling the conditional relationship between sex given gender and other demographics or gender given sex and other demographics. 

There is also other auxiliary information that is available, such as voice tone in a telephone interview or facial recognition software. These should not be used to infer gender unless directly related to the outcome of interest (such as perceived gender discrimination). These systems are complex and traumatic \parencite{Ahmed2017}, are frequently trans exclusive \parencite{Keyes2018Misgendering, lagos2019hearing}, or have racially unbalanced error rates \parencite{buolamwini2018gender}.

\subsubsection*{Impute all non-male respondents as female}

The rationale behind this method is that when we include variables like sex and gender in regression models, we are often most concerned with doing so in a way that best estimates the outcome of interest. For outcomes that vary based on the socially ascribed meanings of sex and gender, what matters most is how these groups are treated differently in society. We therefore might expect non-binary individuals to have outcomes more similar to those of females than those of males, given that they do not benefit from a perceived traditionally masculine gender identity. Therefore, in instances where we cannot adjust for non-binary respondents separately (for reasons of sample size or data security), one reasonable option would be to combine these individuals with those who identify as female.

While intuitive from a sociological perspective, this approaches conflates the constructs of sex and gender even further. It assumes that those who answer the census question on sex are also answering a question on how they are perceived and treated in their society. However, although non-binary respondents do not remain their own group, this is not so much a case of imputing their gender as female, but rather collapsing female and non-binary into one group. Indeed, this variable (in the sample) could be coded as ``male'' and ``not male,'' with those coded as ``not male'' being adjusted to ``female'' in the census. 

\subsubsection*{Remove individuals}

The statistical argument is that the proportion of individuals who respond as non-binary in a survey that does not intentionally recruit from this category is very small (less than 1\% according to various sources; see \cite{meerwijk2017}). Unless this group is very different from those who select male or female, omitting them is unlikely to make a statistical difference to population level estimates (as we will see), but it may make more of a difference when estimating population subgroups.

\subsection*{Comparison of methods} 

To explore the differences between these seven different methods, we conduct a simulation study. Within the simulation study, we manipulated the proportion of non-binary gender respondents who would respond male on a sex question. We also manipulated the difference between genders on the outcome, the response rate of the non-binary respondents and the method of estimation (multilevel regression and poststratification or poststratified weights). A full description of the scenario is available in Appendix A. 

\subsubsection*{Estimating the population mean}

The first statistic that we consider is the population average. Two different methods are used to make this estimate. The first is a weighted average. Raked weights are created using the survey package (\cite{survey1,survey2}). The second is multilevel regression and poststratification (\cite{gelman1997poststratification}) using the poststratification variables as predictors in the model. Other than education and age group, the poststratification variables depend on the method used:  where sex is imputed in the sample, sex is used as a poststratification variable. Where gender is imputed in the population, gender is used as a poststratification variable.

\begin{figure}
    \centering
    \includegraphics[width=\textwidth]{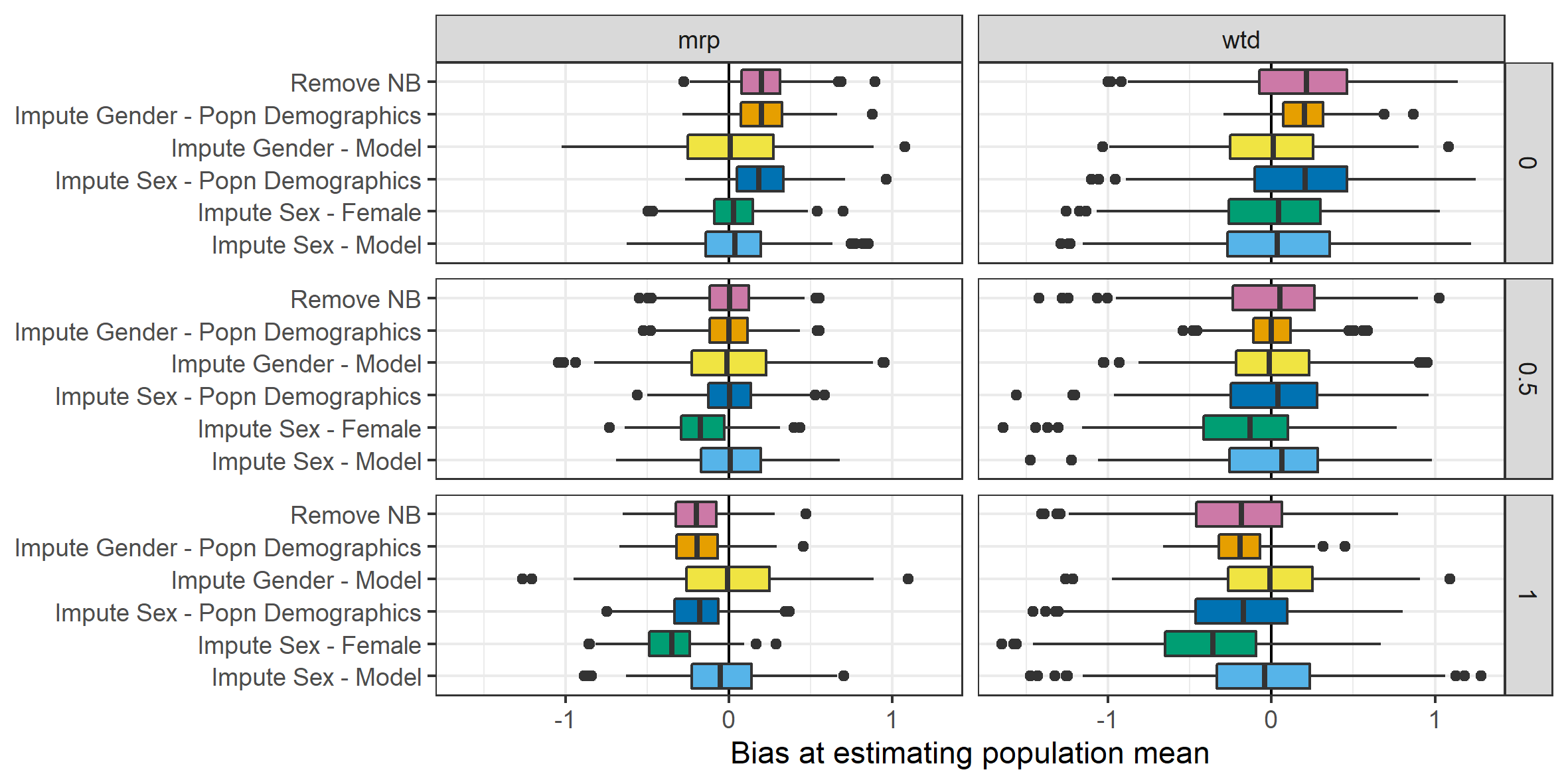}
    \caption{\em Bias of the MRP and weighted estimates from the sample, where all three gender categories have different expected values on the outcome. The colours represent different imputation strategies, and rows represent different hypothetical response by non-binary individuals in a binary sex question (0.5\% and 100\% respond male).}
    \label{fig:simulation1a}
\end{figure}

Figure~\ref{fig:simulation1a} plots the bias calculated over 500 iterations of samples.\footnote{On one iteration, there was an error because there were zero non-binary individuals in the simulated data, which meant that weights could not be created. This iteration was removed from analysis.} Simulated from each population type. We focus on the case where male, female and non-binary genders all have a different expected values (results for other conditions are available in the appendix). The two methods that use a model to impute the gender/sex produce, on average, unbiased estimates of the population mean. Using population demographics to impute gender works well when non-binary individuals are no more likely to choose male than female on a census question (middle row), but produces some bias in other scenarios, whereas imputing sex in the sample (because population demographics provide insight into potential response patterns) does not. Removing non-binary individuals and imputing non-binary individuals as female has a similar pattern across the different hypothetical census response questions.

When we consider the width of the estimates (bearing in mind that imputation uncertainty is not incorporated), we see that the weighted estimates confidence intervals tend to be narrower in width than the corresponding MRP based estimates. When considering the weighted estimates, the impute gender methods tend to have the narrowest uncertainty intervals, while imputing sex using a model has the largest interval width. All other methods perform about the same. We see a slightly different pattern in the MRP based estimates, gender imputed with population demographics having a wider variance. Closer investigation suggests that this is due to greater variance in the estimated population totals when using demographics compared to a modeling approach. 

\begin{figure}
    \centering
    \includegraphics[width=\textwidth]{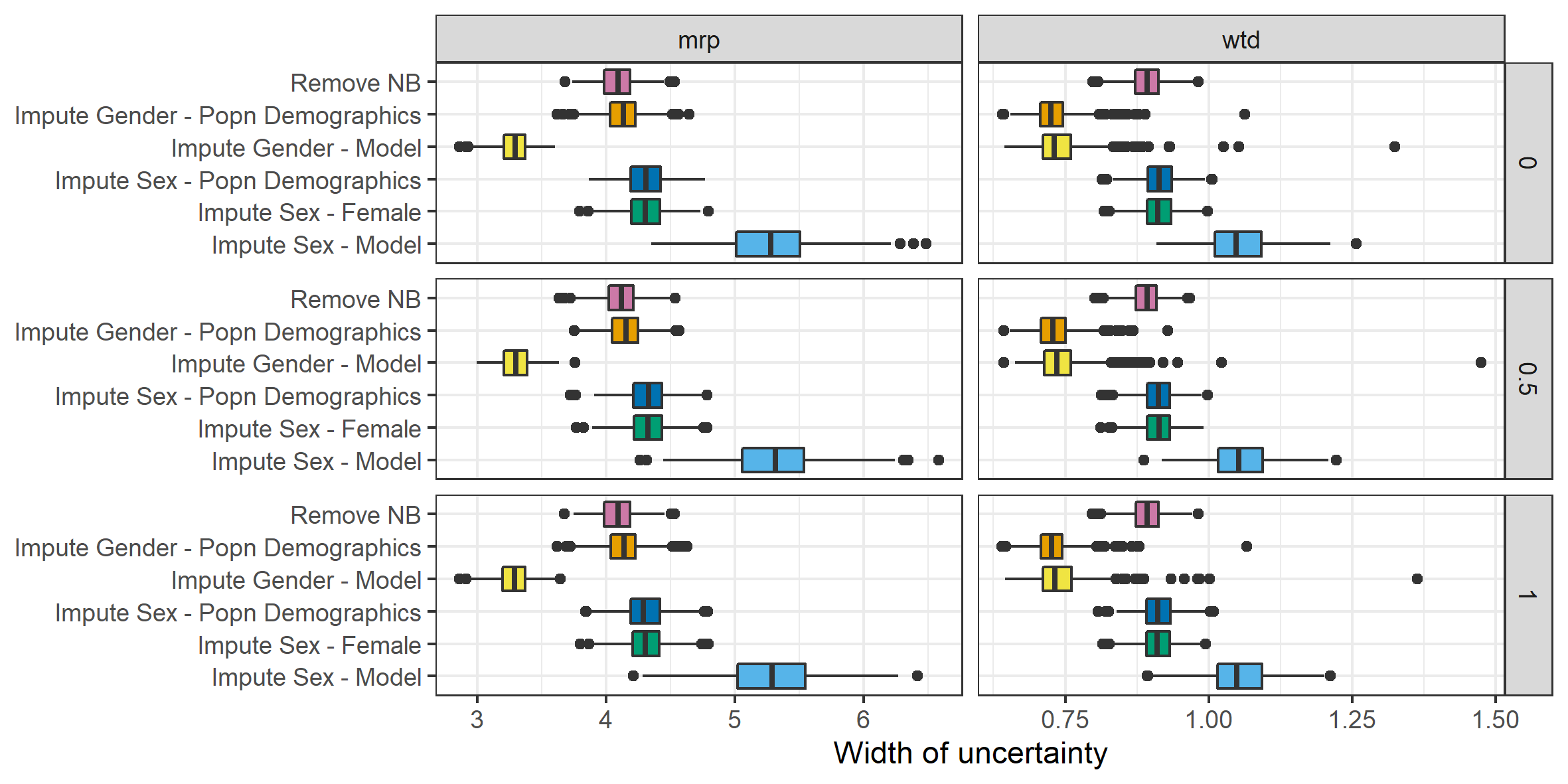}
    \caption{\em Widths of 95\% uncertainty interval for MRP and weighted estimates from the sample, where all three gender categories have different expected values on the outcome. The colours represent different imputation strategies, and rows represent different hypothetical response by non-binary individuals in a binary sex question (0.5\% and 100\% respond male).}
    \label{fig:simulation1b}
\end{figure}

\subsubsection*{Estimating specific sex means}

\begin{figure}
    \centering
    \includegraphics[width=\textwidth]{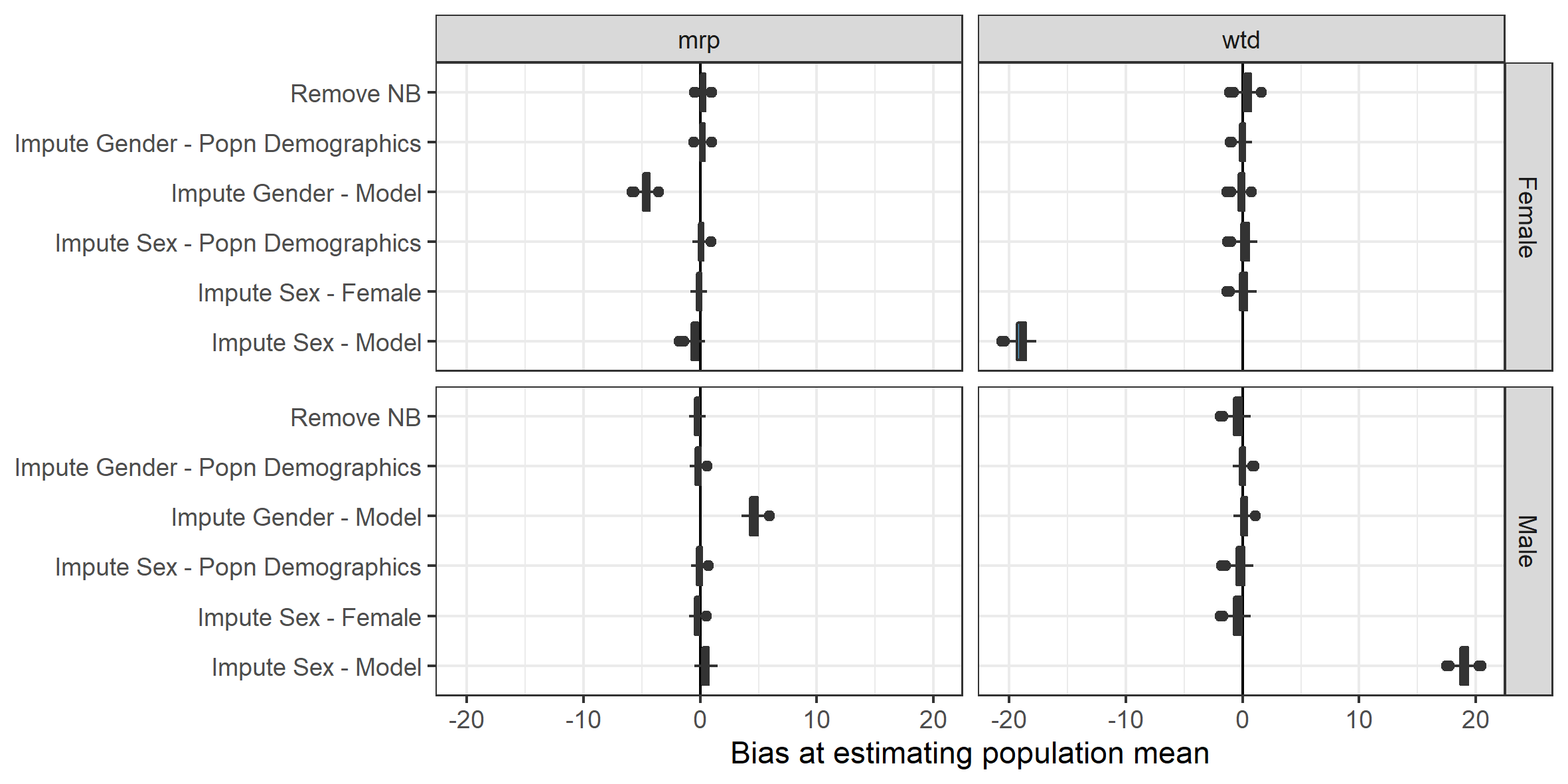}
    \caption{\em Bias of the MRP and weighted estimates from the sample, where all three gender categories have different expected values on the outcome, but no bias on the sex question is expected. The colours represent different imputation strategies, and rows represent the different sex categories.}
    \label{fig:simulation_sex_bias}
\end{figure}

\begin{figure}
    \centering
    \includegraphics[width=\textwidth]{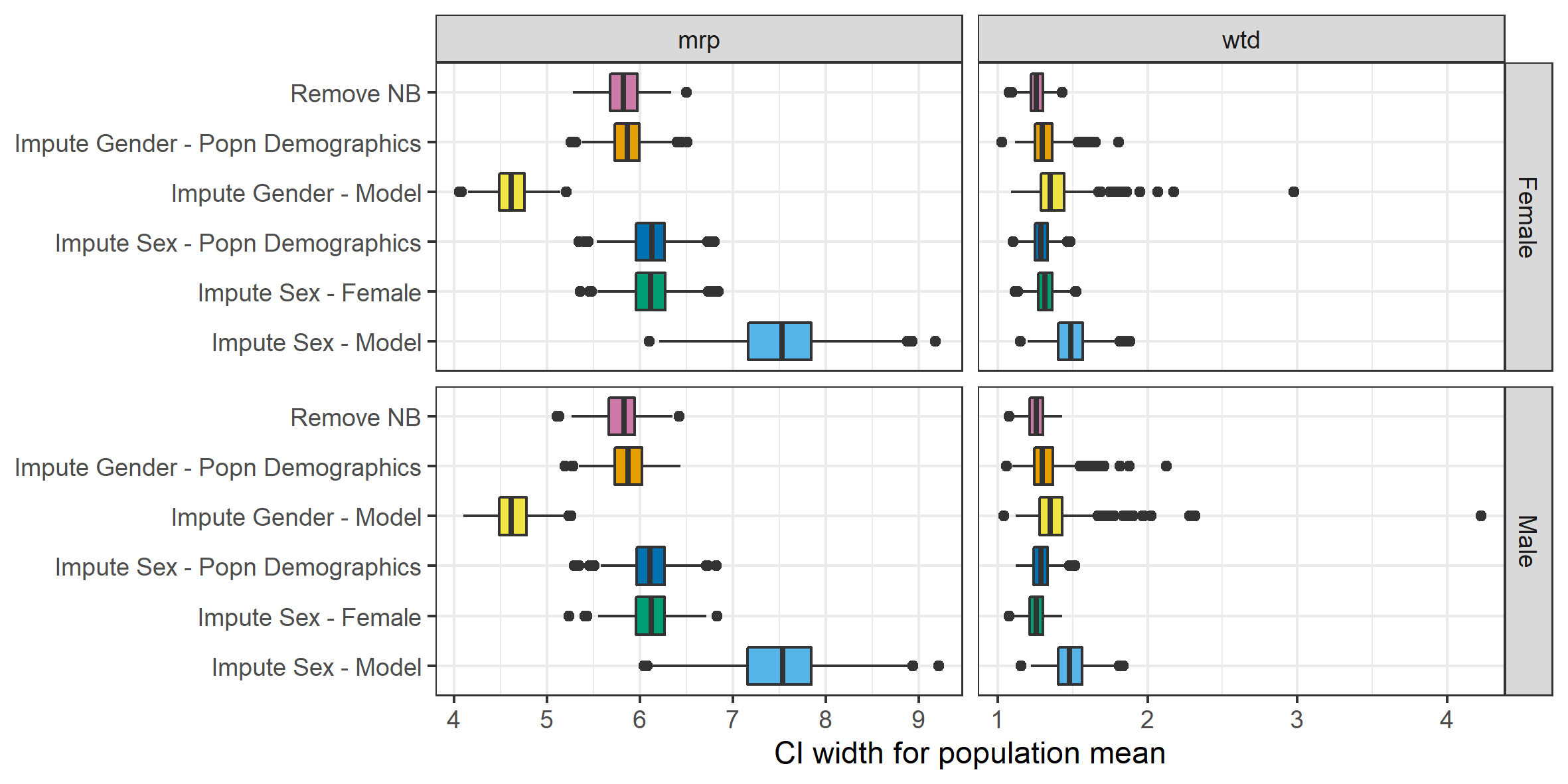}
    \caption{\em Width of 95\% uncertainty of the different sex means, where all three gender categories have different expected values on the outcome, but no bias on the sex question is expected. The colours represent different imputation strategies, and rows represent the different sex categories.}
    \label{fig:simulation_sex_width}
\end{figure}

One reason why we poststratify by sex/gender is to ensure that when we estimate the experiences of different people of different sexes or genders, they are representative of these people. We consider in Figure~\ref{fig:simulation_sex_bias} how the different imputation strategies impact our estimation of male and female sex, using the same simulation strategy as before. For space reasons we present results where male, female, and non-binary individuals were simulated to have different average values of the outcome. As we can see, most imputation methods result in little bias, with two exceptions. The combination of gender imputed with a model and MRP methodology produces some bias, and the combination of sex imputed with a model and a weighted analysis produces considerable bias. These issues are likely due to poor model based predictions. As shown in Figure~\ref{fig:simulation_sex_width}, similar patterns in uncertainty width are observed as in the population total, although imputing gender is associate with slightly more noise when combined with weighted estimation when compared to sex imputation based methods. 

\subsubsection*{Estimating specific gender means}

\begin{figure}
    \centering
    \includegraphics[width=\textwidth]{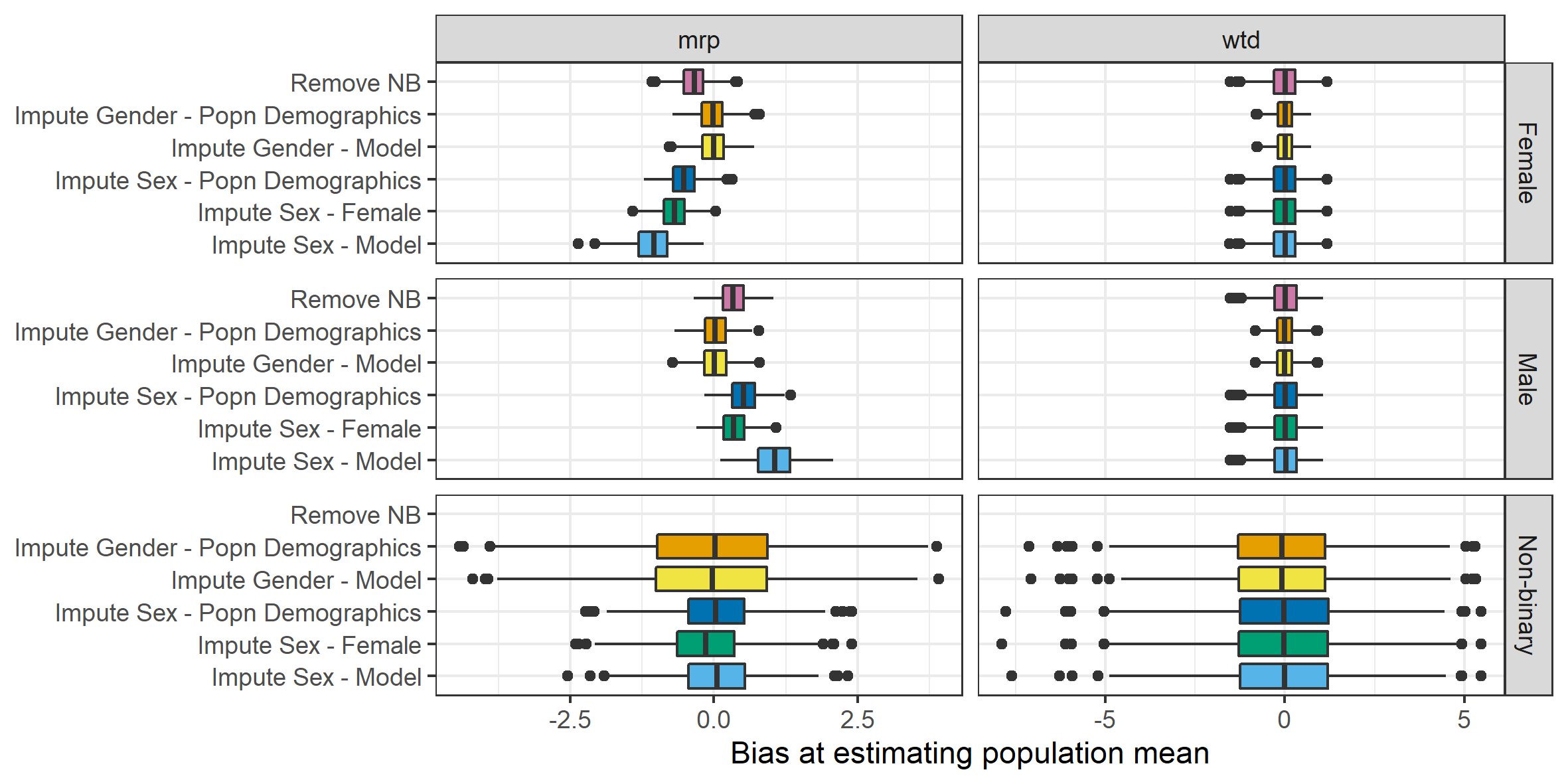}
    \caption{\em Bias of the MRP and weighted estimates from the sample, where all three gender categories have different expected values on the outcome, but no bias on the sex question is expected. The colours represent different imputation strategies, and rows represent the different gender categories.}
    \label{fig:gender_bias}
\end{figure}

\begin{figure}
    \centering
    \includegraphics[width=\textwidth]{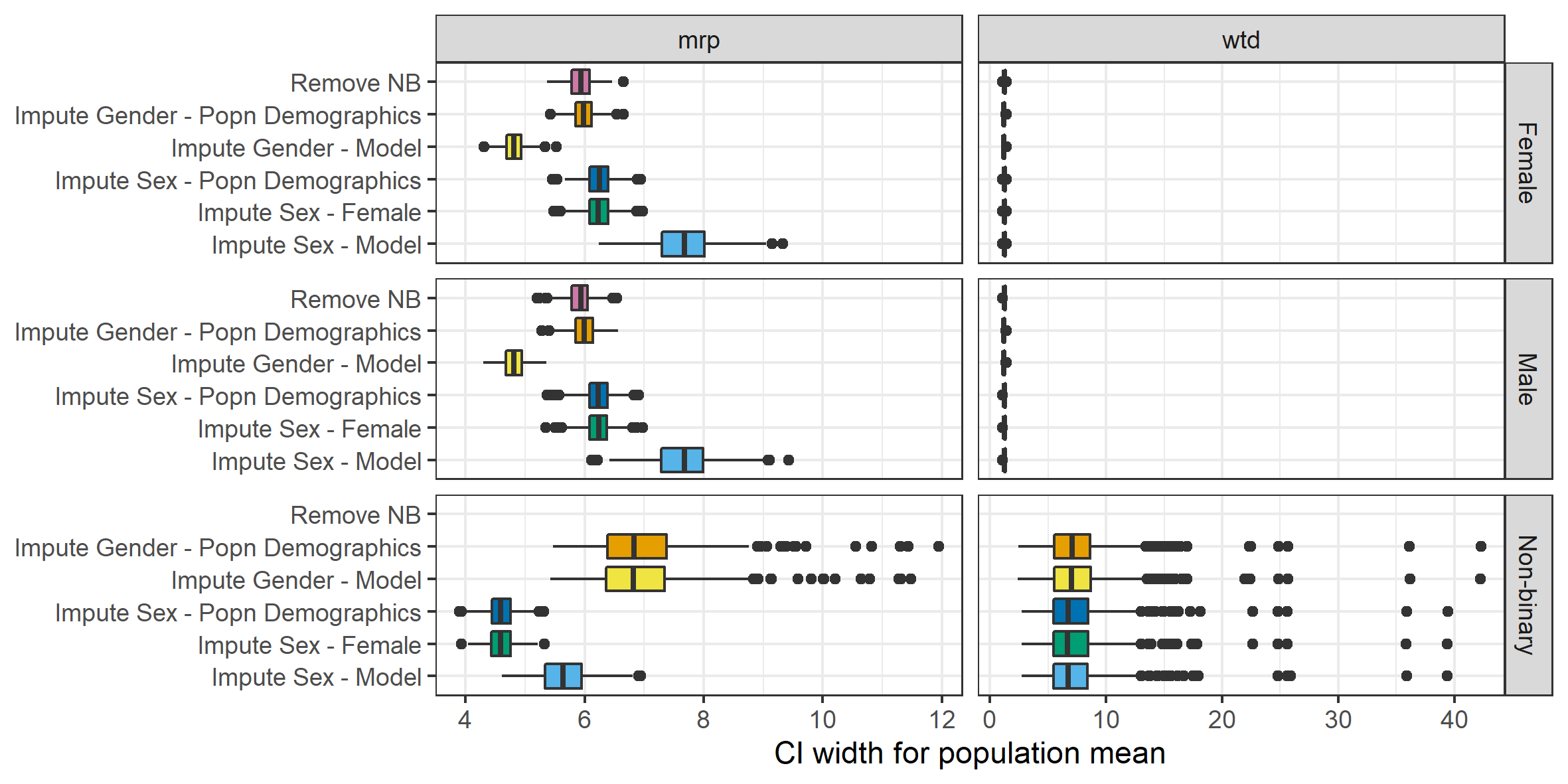}
    \caption{\em Width of 95\% uncertainty interval when estimating the three different gender means in the population. All three gender categories have different expected values on the outcome, but no bias on the sex question is expected. The colours represent different imputation strategies, and rows represent the different gender categories.}
    \label{fig:gender_width}
\end{figure}

Turning out attention to estimating different gender means (Figure~\ref{fig:gender_bias}), we see little difference in method of imputation when considering bias. This makes sense as gender is the measured variable in the sample. Most notable is that the method that removes non-binary individuals means that they do not have an estimate (bottom row, top boxplot). This lack of representation is true for all the statistics with this method, but most salient in this case. Considering the width of uncertainty (Figure~\ref{fig:gender_width}), we see a similar pattern for MRP estimators as previous methods. Notable in weighted estimates the non-binary individuals have less confident estimates, but this reflects the smaller sample sizes and lack of pooling (which is used in an MRP estimate). 

\section{In practice}

The Poverty Tracker was launched in 2012 by Columbia University and the Robin Hood Foundation to track poverty, hardship, and disadvantage in New York City. Since then, the study has enrolled four representative panels of adult New Yorkers and interviewed participants quarterly for up to four years. 

In this study, we considered weighting the random digit dial frames (cellphone and landline) from the 2020 panel of the Poverty Tracker, when it began to include a non-binary gender response category (“Other gender”), to answer the following question: ``Please tell me which gender you identify with: male, female, or something else?'' Before that, gender response categories allowed for only Male and Female answers.\footnote{In this panel, the Poverty Tracker also included two additional frames targeting respondents with Chinese ancestry, which we omit from this analysis for simplicity.}

In this panel, 55.8\% respondents identified as Female ($n=833$), 43.7\% ($n=652$) as Male, and 0.5\% ($n=7$) as Other. Considering the methods outlined above, we decided that only acceptable methods from an ethical perspective that would be suitable in this case were the “Assume known population proportions” and “Remove individuals” methods. The project decided that given the change in response category was specifically due to a desire to increase the inclusiveness of the study, it would not be effective to impute sex in the sample. Although there was some consideration of imputing non-binary respondents as ``not male'' for the purposes of weighting due to an expectation of similarities in disadvantage in the female/non-binary categories, it was ultimately decided that this would also not be suitable. In the rest of this section, we focus on the difference in these two methods when estimating three key outcomes for the survey: the poverty rate, severe material hardship, and whether the respondent had a work-limiting health condition.\footnote{Defined as having a work-limiting health condition or being in self-rated poor health.}

Weights were created using a standard process for the project. An adjustment was made for the number of eligible adults in the household and the frame overlap between cellphone and landline frames. Following this, weights are adjusted to known population totals using raking in the survey package (\cite{survey1}). Adjustment variables include age group, gender (either M/F/NB or M/F depending on the method), race and ethnicity, education level, immigrant status, housing situation, proportion of the year worked, ratio of income to the household specific poverty line, and basic household demographics: number of older adults, working age adults and children. This follows the standard weighting process of previous panels, with the only manipulation being how the gender variable is accounted for.

Given the low population level information in the U.S. of gender demographics, we assumed that the population proportions were consistent with those observed in the sample.\footnote{We tried to benchmark to gender population data in New York City or New York State. The limited source we found from NYC Health Department suggested that the rough estimation of 0.5\% for the non-binary gender group in our sample was reasonable. Please see \url{https://www1.nyc.gov/site/doh/about/press/pr2019/non-binary-gender-category-to-nyc-death-certificates.page} and \url{https://www.health.ny.gov/statistics/brfss/reports/docs/1806_brfss_sogi.pdf} for more information.} Using this ``impute population values'' method, the estimated poverty rate was 23.55\% with a standard error of 1.72\%. When removing individuals who responded Other to the gender question, the results yield a similar poverty rate of 23.65\% with the same standard error. The difference between these two estimates is tiny compared to the standard error, but in one version non-binary individuals are represented in the estimate, whereas in the other they are not.

We further looked at the subgroup poverty estimates; again, they were largely similar between the two methods; see Table~\ref{tab:pt_ests} for details. One major difference between these methods was that when non-binary individuals are removed from the data before poststratification, estimates for the non-binary subgroup cannot be made, whereas they can in the alternate method. We chose not to release the individual estimates for the non-binary individuals due to the small cell size ($n=7$) and data privacy concerns. However the possibility of producing an estimate for non-binary individuals enables greater understanding of the experiences of this gender group. Even if results are not publicly reported, they can still be used internally to plan future survey panels as well as assess the impact of policy on this particular group.

\begin{table}
    \centering
    \begin{tabular}{ M{2.4cm}M{2.7cm}M{2.2cm}M{2.2cm}M{2.2cm}M{2.2cm}}
     \multirow{3}{*}{Estimate}&\multirow{3}{*}{Condition}&
      \multicolumn{3}{c}{Gender group} &
       Overall\\
     && Female & Male & Non-binary & \multirow{2}{*}{mean (s.e.)}\\
     & & mean (s.e.) & mean (s.e.)& mean (s.e.) & \\
     \hline
     \multirow{4}{2.4cm}{Poverty}& Assume known population proportions & 26.79\% (2.46\%) & 20.82\% (2.42\%) & Redacted & 23.55\% (1.72\%) \\[1cm]
     & Remove individuals & 26.80\% (2.46\%) & 20.85\% (2.42\%) & NA & 23.65\% (1.72\%)\\
    \hline
    \multirow{4}{2.4cm}{Severe Hardship} &   Assume known population proportions & 37.57\% (2.82\%) & 30.55\% (2.71\%) & Redacted & 33.62\% (2.32\%) \\[1cm]
     & Remove individuals & 37.62\% (2.82\%) & 30.66\% (2.71\%) & NA & 33.93\% (1.96\%)\\
    \hline
    \multirow{4}{2.4cm}{Limiting Health} &Assume known population proportions & 21.65\% (2.28\%) & 15.40\% (2.03\%) & Redacted & 18.44\% (1.53\%)\\[1cm]
     & Remove individuals & 21.57\% (2.28\%) & 15.37\% (2.03\%) & NA & 18.28\% (1.52\%) 
    \end{tabular}
    \caption{\em Comparison between two methods estimating poverty, severe hardship, and work-limiting health conditions. We display to an over-precise two decimal places to show how tiny the differences are when estimating these aggregate proportions.}
    \label{tab:pt_ests}
\end{table}

Turning our attention to other key outcomes in the survey we see similar findings. Using weights created with a gender imputed population, the results yielded a severe hardship rate estimate of 32.76\% with a standard error of 2.0\%. Using the weights created by removing individuals who responded in Other, the results were similar with a severe hardship estimate of 32.74\% with the same standard error. Differences were also tiny when estimating the proportion of New Yorkers experiencing limiting health conditions.

Comparing the subgroup estimates, we see similar findings. However, two features are salient. Firstly, although largely similar, it is only possible to produce estimates for the non-binary individuals in the method where they are not removed. These estimates are redacted due to small sample size in this application, but this highlights that the method of removing individuals does not allow for representation of non-binary individuals in any weighted estimate. There are significant gender differences on all three outcomes with women reporting higher rates of poverty, severe hardship, and work-limiting health conditions (6\%, 6\%, and 7\%, respectively). This suggests that there are gender differences present, and perhaps dedicated efforts to increase the proportion of non-binary individuals would be merited. 

Overall, the results suggest that when it comes to estimating specific outcomes for gender groups, the method of choice makes little difference: the differences are in the hundredth of a percentage point for population proportions. The lack of statistical differences means that we can make decisions from an ethical framework. Differences between genders on our three outcomes provide support that there is substantial gender-related heterogeneity. This is  why we chose to use a method imputing gender in the population as it enables us to create an estimate for all three gender responses. 

\section{Looking forward}

Measurement is central to science and statistics and represents a particular challenge to survey researchers and social scientists because the constructs that we measure are changing in both importance and definition over time. This means that an appropriate measurement of a construct today might not be an appropriate measure of the construct tomorrow. Indeed, measurement in the social sciences reflects the sociological emphasis that is placed on the underlying construct. This is a challenge faced when considering the construct of race/ethnicity, but this challenge is also faced when considering sex/gender. Our challenge increases when we consider that we are not simply moving to a more diverse way of coding sex, but instead a recognition that the construct of gender, while the same as sex assigned at birth for many, is a different construct for others. This distinction led us to frame our methods in terms of imputing one construct from the other. 

This manuscript grapples with the complexities of moving from measuring sex to measuring gender in social surveys. What it does not do, however, is make broad recommendations for a one best way to measure sex or gender or a one best technique to account for measuring gender in a survey when the population measures sex. Instead we try to consider the ethical and statistical implications of a variety of different approaches.

Constructing our argument in this way is necessary, as there is no single good solution that can be applied to all situations. Instead it is important to recognize that there is a compromise between ethical concerns, statistical concerns, and the most appropriate decision will be reflective of this. That said, we have argued that first and foremost in this decision should be respect and consideration for the survey respondent, followed by the ease of describing the statistical method to non-technical respondents and concerns surrounding fair representation and statistical bias.

Enumerating the potential options to Scenario 1 in Table~\ref{tab:pot_options_scenario1}, statistical and ethical concerns intersect, which implies that both facets need to be considered. While specific to the challenges of measuring sex and gender, our review of these approaches, with their various advantages and tradeoffs, may be useful in grappling with the measurement of other social constructs as well. Our hope is that this is a useful resource to guide decision making for survey statisticians and survey administrators alike. 

\printbibliography

\section{Simulation study method} 

Instead of focusing on the differences between sex and gender as constructs, we focus on the difference in respondent perception when answering a male-female response question, ``What is your sex,'' versus a male-female-non-binary response to the question, ``What gender do you identify as?''  While sex and gender are different constructs, from the perspective of matching, the difference in responses between these two questions is how they are interpreted by the respondent. 

To this end we create a simulation strategy where we simulate a population where proportions $(p_m,p_f,p_o)$ respond male/female/non-binary when asked what gender they identify with. We then assume that a portion who respond male/female to this question also respond male/female when asked what sex they are. For those who respond non-binary to the first question, we simulate that they respond male to the second question at 0\% (no one), 50\% (equal chance) and 100\% (everyone). A summary of the response patterns is given in Table~\ref{tab:response_pattern}. Studies investigating the difference in response to these questions have estimated that less than 1\% of the population would respond as non-binary when given the option \parencite{meerwijk2017}. However, this would depend on both the question framing (see the much higher proportions when gender is expressed as a continuum) and option availability (e.g., the Australian census required respondents to request the non-binary category rather than simply offering it). In our simulation study we use population gender proportions of 49\% female, 49\% male, and 2\% non-binary.

For each scenario we simulate a sample of size 500. We then imagine potential response strategies to a question eliciting binary sex given responses to gender. We assume that a small proportion (2\%) of those who respond female to the gender question would respond male to a sex question, and a similar proportion (2\%) of those who respond male to the gender question would respond female to the sex question. We then vary the proportion of those who respond non-binary to the gender question who would respond male to a sex question from 0 to 100\%; see Table~\ref{tab:response_pattern} for enumeration of the possibilities.

We consider an outcome variable 
$y_i \sim \operatorname{normal}(\mu_i, 1)$, where
\begin{equation*}
         \mu_i = \Bigg\{ \begin{array}{ll}
         \mu_\text{m} & \textrm{if gender[i] is male}\\
         \mu_\text{f} & \textrm{if gender[i] is female}\\
         \mu_\text{nb} & \textrm{if gender[i] is non-binary.}\\
    \end{array} 
\end{equation*}
We then simulate four different hypothetical conditions:
\begin{enumerate}
    \item There are no differences between gender in the outcome, $\mu_\text{m} = \mu_\text{f} = \mu_\text{nb}$.
    \item Those who identify as non-binary have different outcomes when compared to those who identify as male and female, $\mu_\text{m} = \mu_\text{f}; \mu_\text{m} \neq \mu_\text{nb}$.
    \item Those who identify as female or non-binary are different in outcome from those who identify as male, $\mu_\text{m} \neq \mu_\text{nb}; \mu_\text{nb} = \mu_\text{f}$.
    \item There are different outcomes for respondents who identify as different genders,  $\mu_\text{m} \neq \mu_\text{f} \neq \mu_\text{nb}$.
\end{enumerate}
We then compare seven potential options for adjusting male/female/non-binary measurements in the sample to a male/female measurement in the population:
\begin{enumerate}
    \item Impute all who identify as non-binary as male or female with a 50/50 split in the sample.
    \item Impute all who identify as non-binary as female.
    \item Impute sex by gender information using a model in the sample:
    \begin{enumerate}
        \item Simulate a best-case model imputing male as male/female in the correct proportions and vice versa, and non-binary relative to the proportion who choose to respond M/F.
        \item Simulate a worst-case model imputing male as male/female in the opposite proportions and vice versa, and non-binary opposite to the proportion who choose to respond M/F.
    \end{enumerate}
      \item Impute  $\text{gender}\times\text{sex}$  information using a model in the population:
    \begin{enumerate}
        \item Simulate a best case model (imputes male, female, and non-binary counts in the correct proportions).
        \item Simulate a worst case model (imputes male, female, and non-binary counts in the opposite proportions).
    \end{enumerate}
    \item Remove those who respond as non-binary and map male gender to male sex and female gender to female sex.
\end{enumerate}

Then, either we use the imputed sex counts to create simple poststratification weights for the sample and then use a weighted average of the sample to estimate the population mean, or we use the imputed gender counts to create simple poststratification weights for the sample and then use a weighted average of the sample to estimate the population mean.

\begin{table}
    \centering
    \begin{tabular}{cc c c}
        Condition & Male $\mu$ & Female $\mu$ & non-binary $\mu$ \\
        \hline
        All same &  0 & 0 &0 \\
        Male, female same & 10 & 10 & 0 \\
        Female, non-binary same & 10 & 0 & 0 \\
        All different & 10 & $-10$ & 0 
    \end{tabular}
    \caption{\em Simulations of  $\mu$ for each gender category for each of the four conditions for the first simulation scenario. The standard deviation for the outcome is 4, so these represent large effect sizes.}
    \label{tab:effect_size_lg}
\end{table}

\begin{table}
    \centering
    \begin{tabular}{l c c}
     & Male sex & Female sex \\
     \hline
     Male gender & 48\% & 1\% \\
     Female gender & 1\% & 48\% \\     Non-binary gender & $p \times 2\%$ & $p \times 2\%$ 
    \end{tabular}
    \caption{\em Population distribution of male, female, and non-binary genders relative to a binary sex variable in a simulation study. The proportion of respondents with a non-binary gender who select male or female sex is varied from all male to all female in the simulation.}
    \label{tab:response_pattern}
\end{table}

Two other demographic variables, representing age group (3 levels) and education (3 levels), were also included to represent the minimum complexity expected for a poststratification type analysis.

\section{Simulation study results} 

\subsection{All gender groups differ on outcome }

\subsubsection{Population estimation}

\begin{figure}[H]
    \centering
    \includegraphics[width=\textwidth]{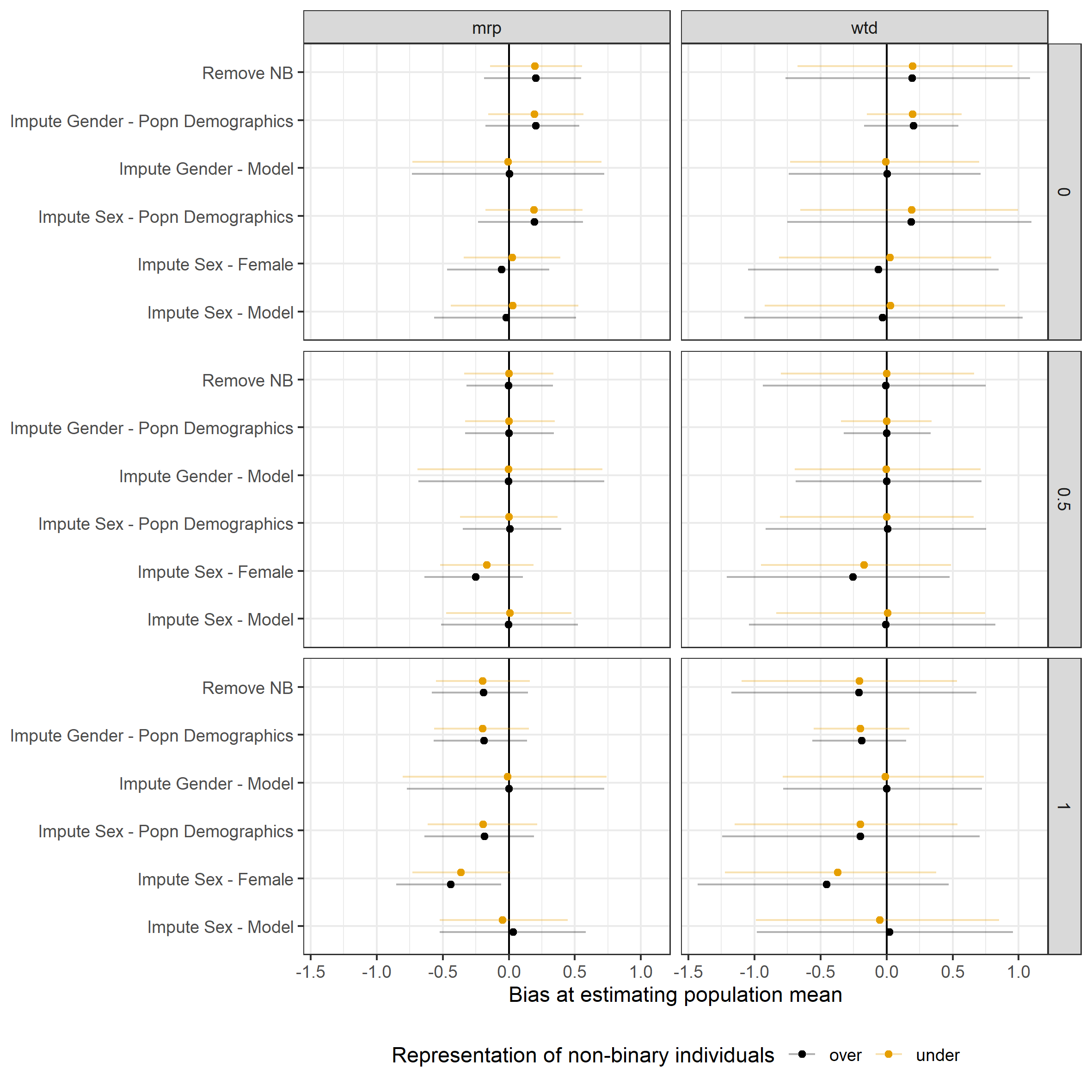}
    \caption{\em Bias of the MRP and weighted estimates from the sample, where all three gender categories have different expected values on the outcome. The three row facets represent the different proportions of non-binary individuals who respond Male in response to a sex question. The two columns represent the analysis method (MRP and poststratification weights). Points represent the mean bias, with lines representing the 95\% quantiles across the 500 simulated data sets. Colour represents whether non-binary individuals were over or under represented in the data (under represented condition reported in main text).}
\end{figure}

\begin{figure}[H]
    \centering
    \includegraphics[width=\textwidth]{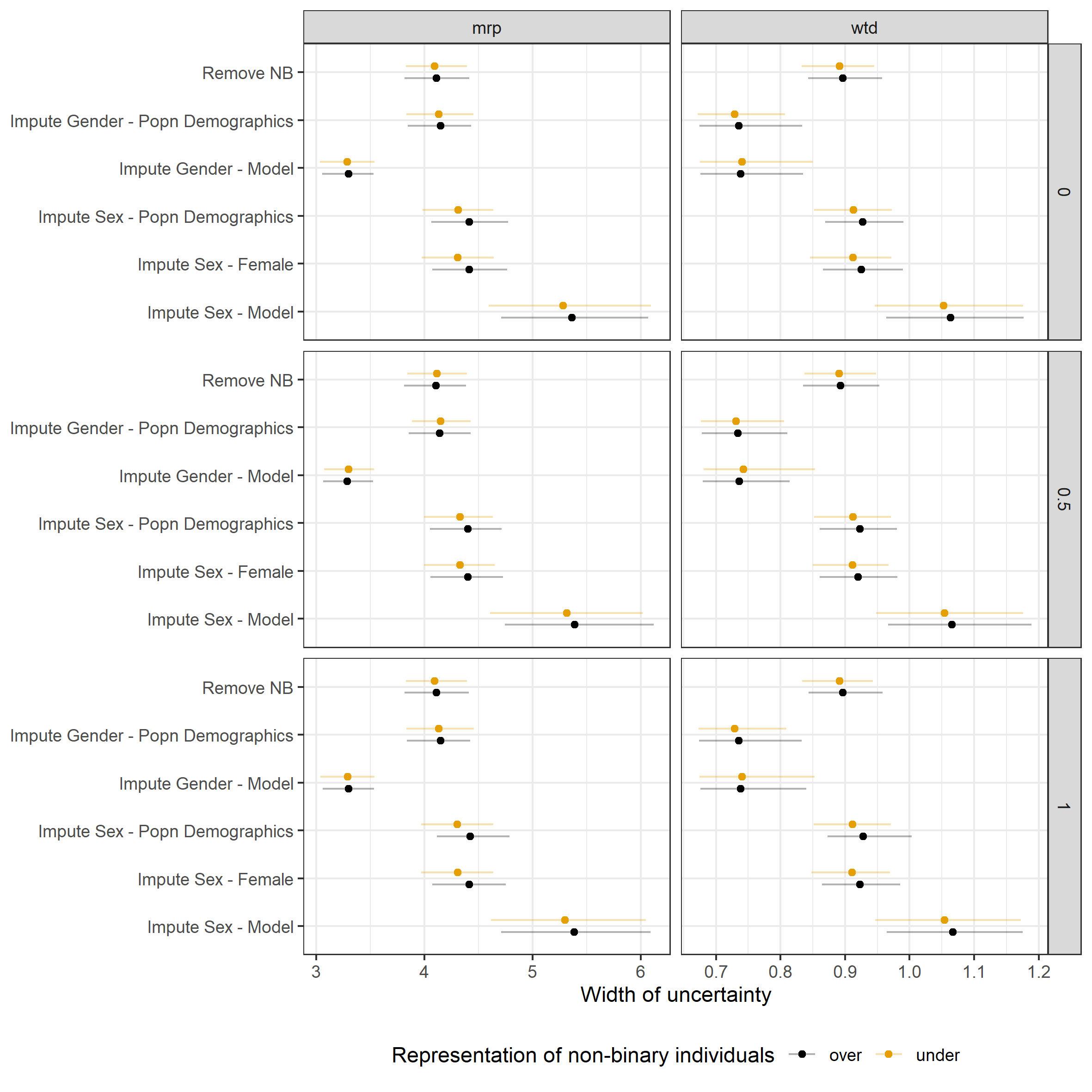}
    \caption{\em Width of uncertainty interval of the MRP and weighted estimates from the sample, where all three gender categories have different expected values on the outcome. The three row facets represent the different proportions of non-binary individuals who respond Male in response to a sex question. The two columns represent the analysis method (MRP and poststratification weights). Points represent the mean bias, with lines representing the 95\% quantiles across the 500 simulated data sets. Colour represents whether non-binary individuals were over or under represented in the data (under represented condition reported in main text).}
\end{figure}

\subsubsection{Sex demographic estimation}

\begin{figure}[H]
    \centering
    \includegraphics[width=\textwidth]{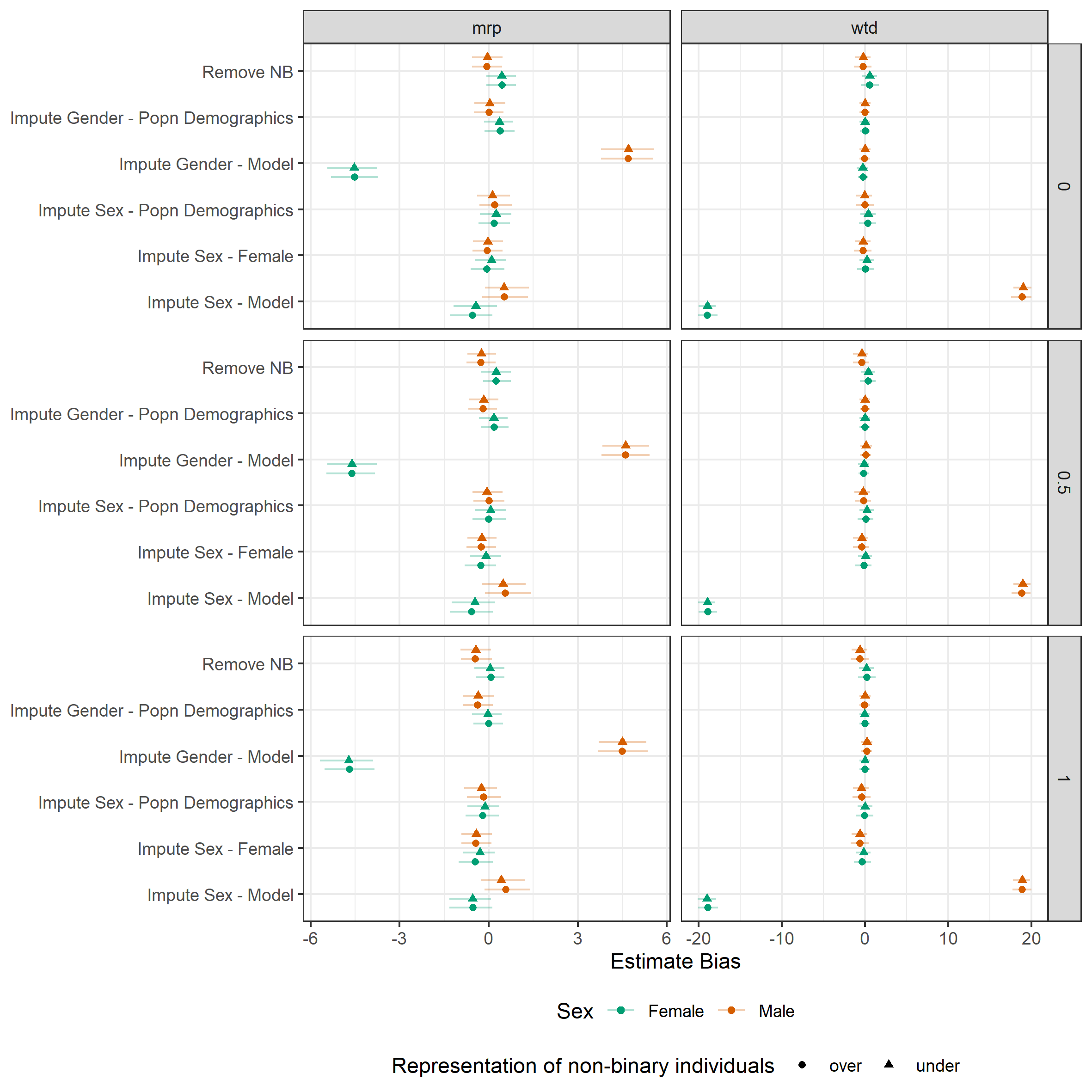}
    \caption{\em Bias of estimates of different sex demographics (M/F) when compared to population truth, where all three gender categories have different expected values on the outcome. The three row facets represent the different proportions of non-binary individuals who respond Male in response to a sex question. The two columns represent the analysis method (MRP and poststratification weights). Points represent the mean bias, with lines representing the 95\% quantiles across the 500 simulated data sets. Colour represents the sex demographic being predicted, while shape represents whether non-binary individuals were over or under represented in the data (under represented condition reported in main text). The findings are similar to the main text, with the model based methods demonstrating worst bias, with imputing sex worse for weighted methods, and imputing gender worse for MRP-based methods. Pattern largely remains the same as that reported in text (middle row). }
\end{figure}

\begin{figure}[H]
    \centering
    \includegraphics[width=\textwidth]{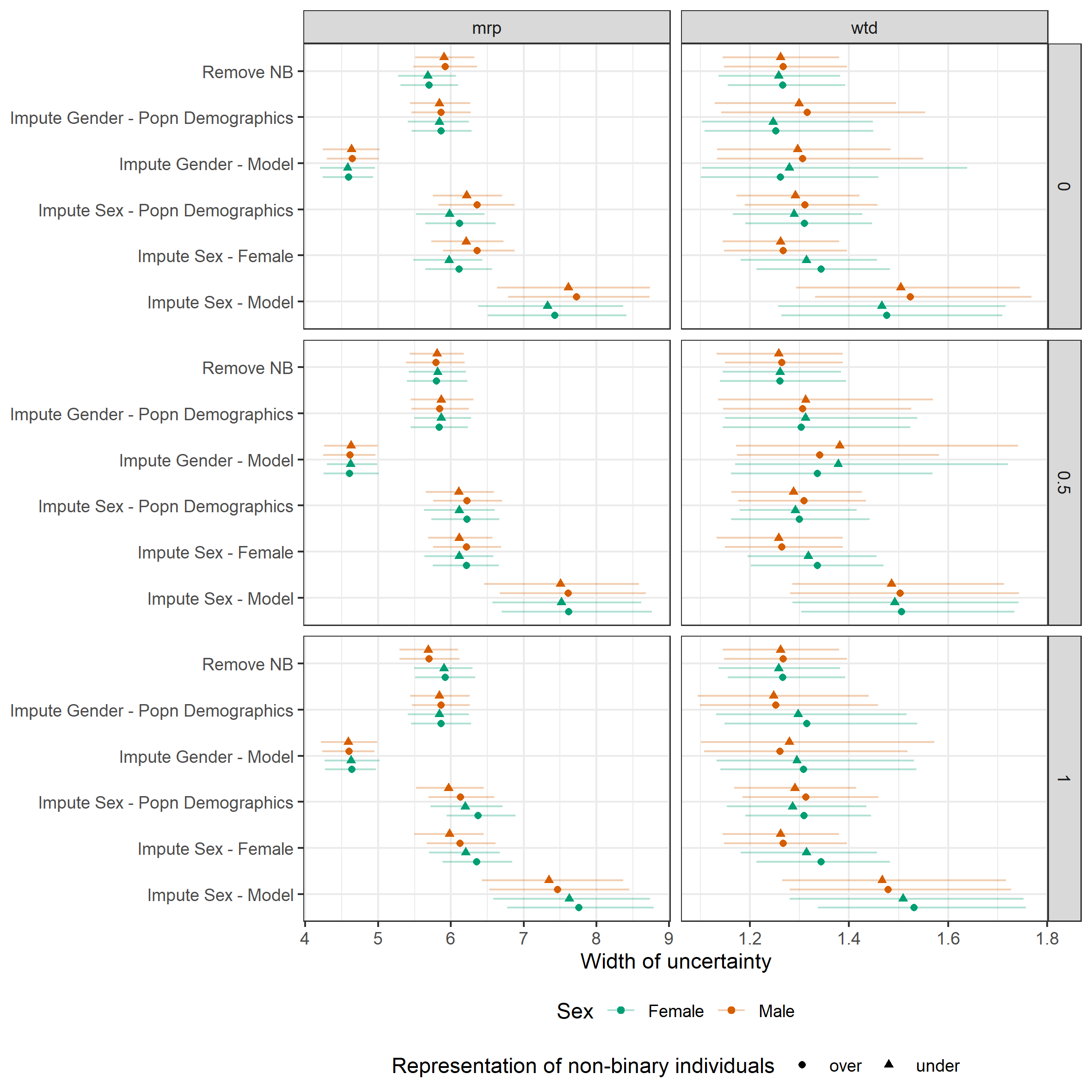}
    \caption{\em Width of uncertainty intervals of estimates of different sex demographics (M/F) when compared to population truth, where all three gender categories have different expected values on the outcome. The three row facets represent the different proportions of non-binary individuals who respond Male in response to a sex question. The two columns represent the analysis method (MRP and poststratification weights). Points represent the mean bias, with lines representing the 95\% quantiles across the 500 simulated data sets. Colour represents the sex demographic being predicted, while shape represents whether non-binary individuals were over or under represented in the data (under represented condition reported in main text). Pattern largely remains the same as that reported in text (middle row). }
\end{figure}

\subsubsection{Gender demographic estimation}

\begin{figure}[H]
    \centering
    \includegraphics[width=\textwidth]{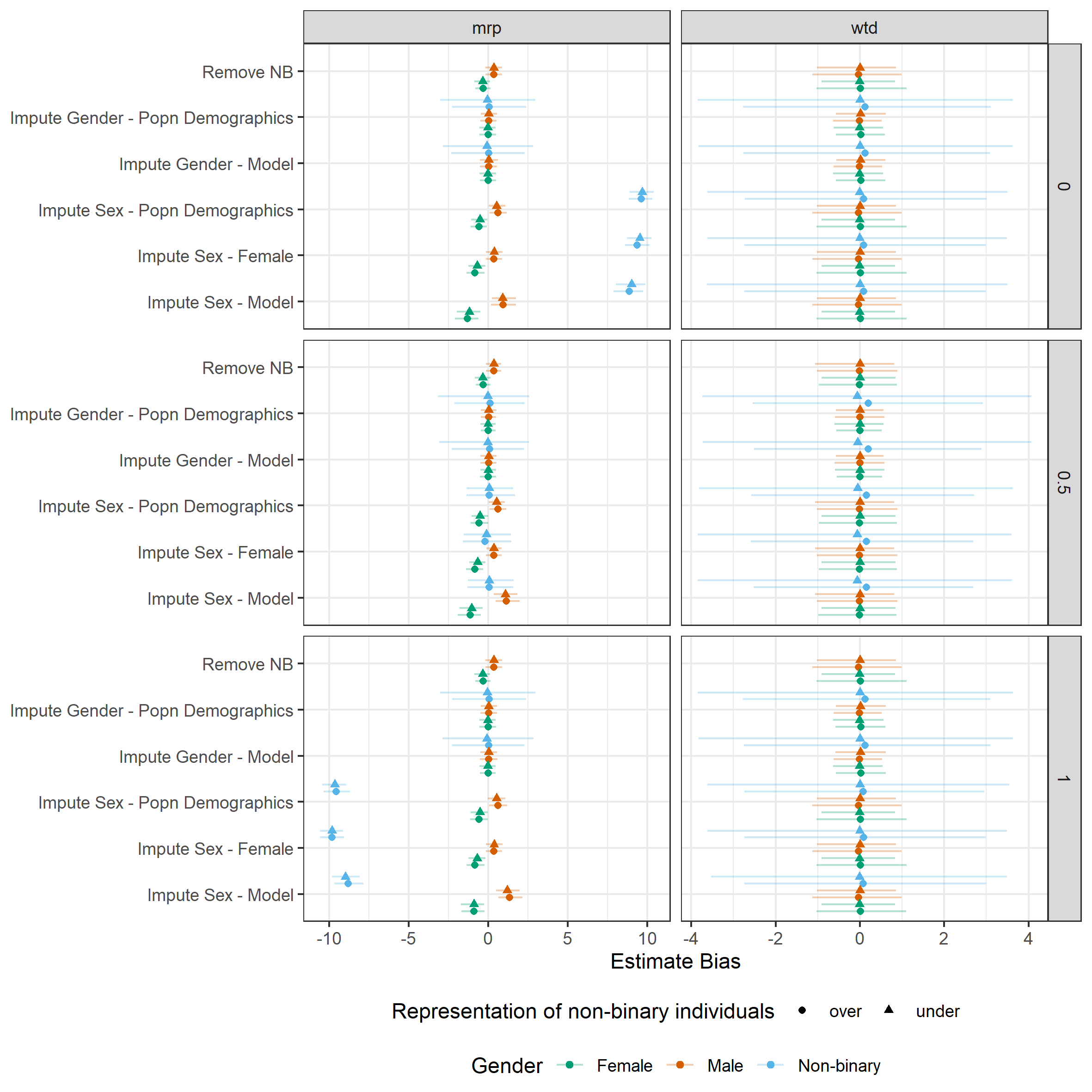}
    \caption{\em Bias of estimates of different gender demographics (M/F/NB) when compared to population truth, where all three gender categories have different expected values on the outcome. The three row facets represent the different proportions of non-binary individuals who respond Male in response to a sex question. The two columns represent the analysis method (MRP and poststratification weights). Points represent the mean bias, with lines representing the 95\% quantiles across the 500 simulated data sets. Colour represents the gender demographic being predicted, while shape represents whether non-binary individuals were over or under represented in the data (under represented condition reported in main text). The findings demonstrate the biased responses have a large impact of the efficacy of MRP sex imputation based methods, which was not reported in the main text (middle row). }
\end{figure}

\begin{figure}[H]
    \centering
    \includegraphics[width=\textwidth]{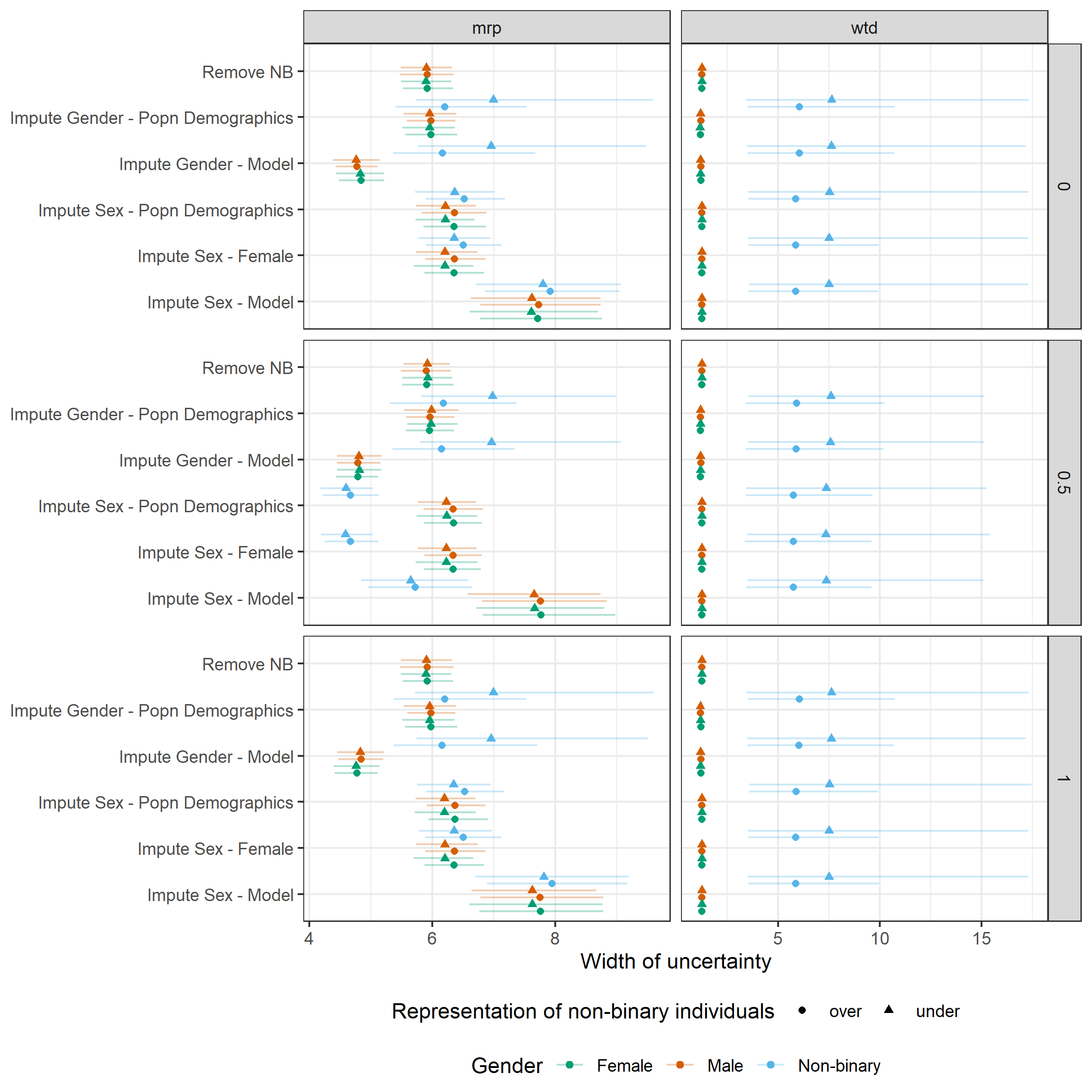}
    \caption{\em Width of uncertainty intervals of estimates of different gender demographics (M/F/NB) when compared to population truth, where all three gender categories have different expected values on the outcome. The three row facets represent the different proportions of non-binary individuals who respond Male in response to a sex question. The two columns represent the analysis method (MRP and poststratification weights). Points represent the mean bias, with lines representing the 95\% quantiles across the 500 simulated data sets. Colour represents the gender demographic being predicted, while shape represents whether non-binary individuals were over or under represented in the data (under represented condition reported in main text). Pattern largely remains the same as that reported in text (middle row). }
\end{figure}

\subsection{Non-binary individuals simulated as responding differently to male/female respondents.}

\subsubsection{Population estimation}

\begin{figure}[H]
    \centering
    \includegraphics[width=\textwidth]{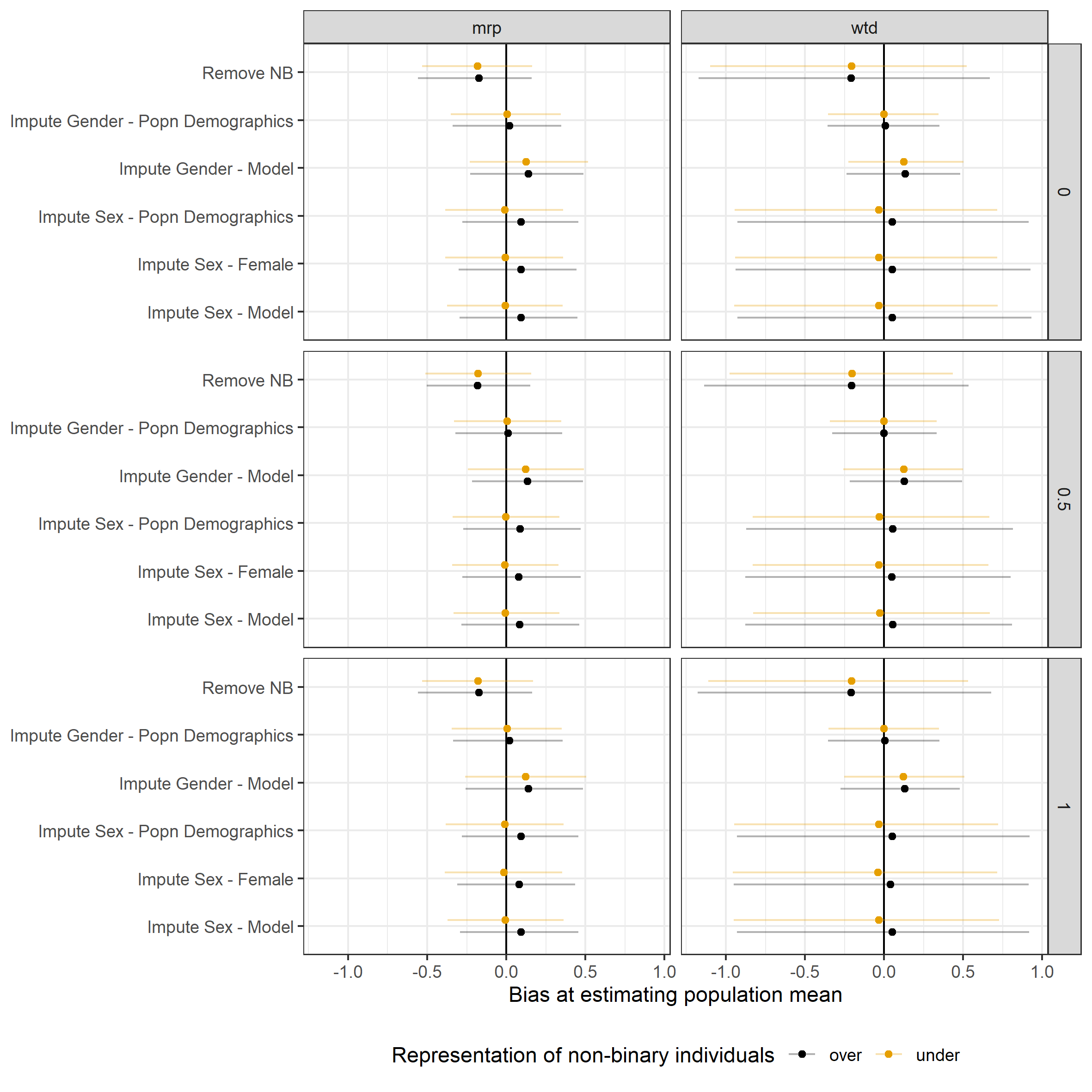}
    \caption{\em Bias of the MRP and weighted estimates from the sample, where non-binary genders have different expected values on the outcome. The three row facets represent the different proportions of non-binary individuals who respond Male in response to a sex question. The two columns represent the analysis method (MRP and poststratification weights). Points represent the mean bias, with lines representing the 95\% quantiles across the 500 simulated data sets. Colour represents whether non-binary individuals were over or under represented in the data. This condition was not reported in the main text, but demonstrates overall unbiasedness of most methods in this condition. Removing non-binary individuals and imputing gender with a model have some small bias. The reason for the former is obvious, the reason for the latter is likely due to the difficulty of obtaining good gender predictions from the other demographics included in the simulation.}
\end{figure}

\begin{figure}[H]
    \centering
    \includegraphics[width=\textwidth]{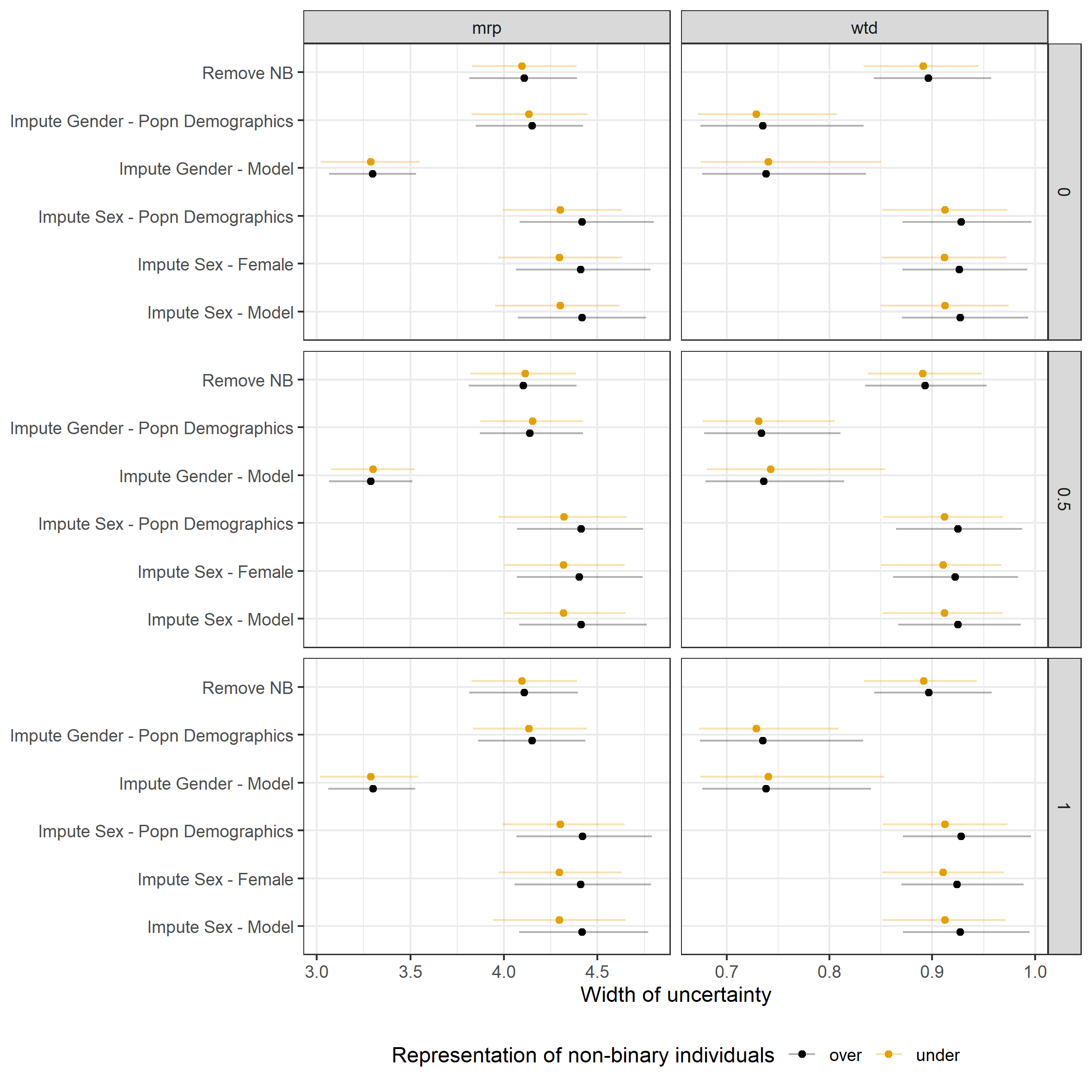}
    \caption{\em Width of uncertainty interval of the MRP and weighted estimates from the sample, where non-binary genders have different expected values on the outcome. The three row facets represent the different proportions of non-binary individuals who respond Male in response to a sex question. The two columns represent the analysis method (MRP and poststratification weights). Points represent the mean bias, with lines representing the 95\% quantiles across the 500 simulated data sets. Colour represents whether non-binary individuals were over or under represented in the data. This condition was not reported in the main text, but represents similar findings. The main differences is the similarity between the imputing sex using a model and the other sex imputation methods. }
\end{figure}

\subsubsection{Sex demographic estimation}

\begin{figure}[H]
    \centering
    \includegraphics[width=\textwidth]{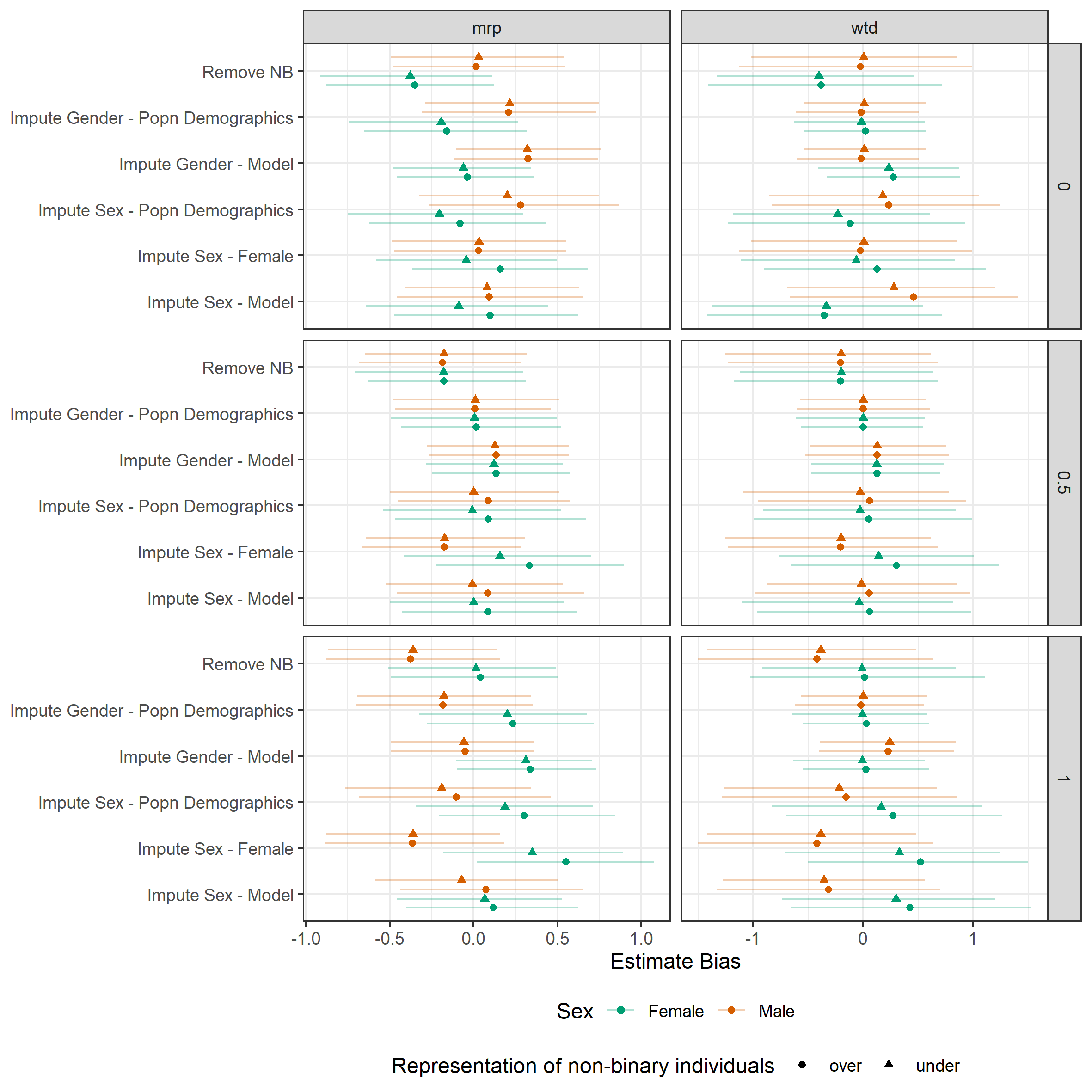}
    \caption{\em Bias of estimates of different sex demographics (M/F) when compared to population truth, where non-binary genders have different expected values on the outcome. The three row facets represent the different proportions of non-binary individuals who respond Male in response to a sex question. The two columns represent the analysis method (MRP and poststratification weights). Points represent the mean bias, with lines representing the 95\% quantiles across the 500 simulated data sets. Colour represents the sex demographic being predicted, while shape represents whether non-binary individuals were over or under represented in the data. These findings were not reported in the main text. The findings suggest greater noise between the methods and sex biases than the main text, but with no clear best method.}
\end{figure}

\begin{figure}[H]
    \centering
    \includegraphics[width=\textwidth]{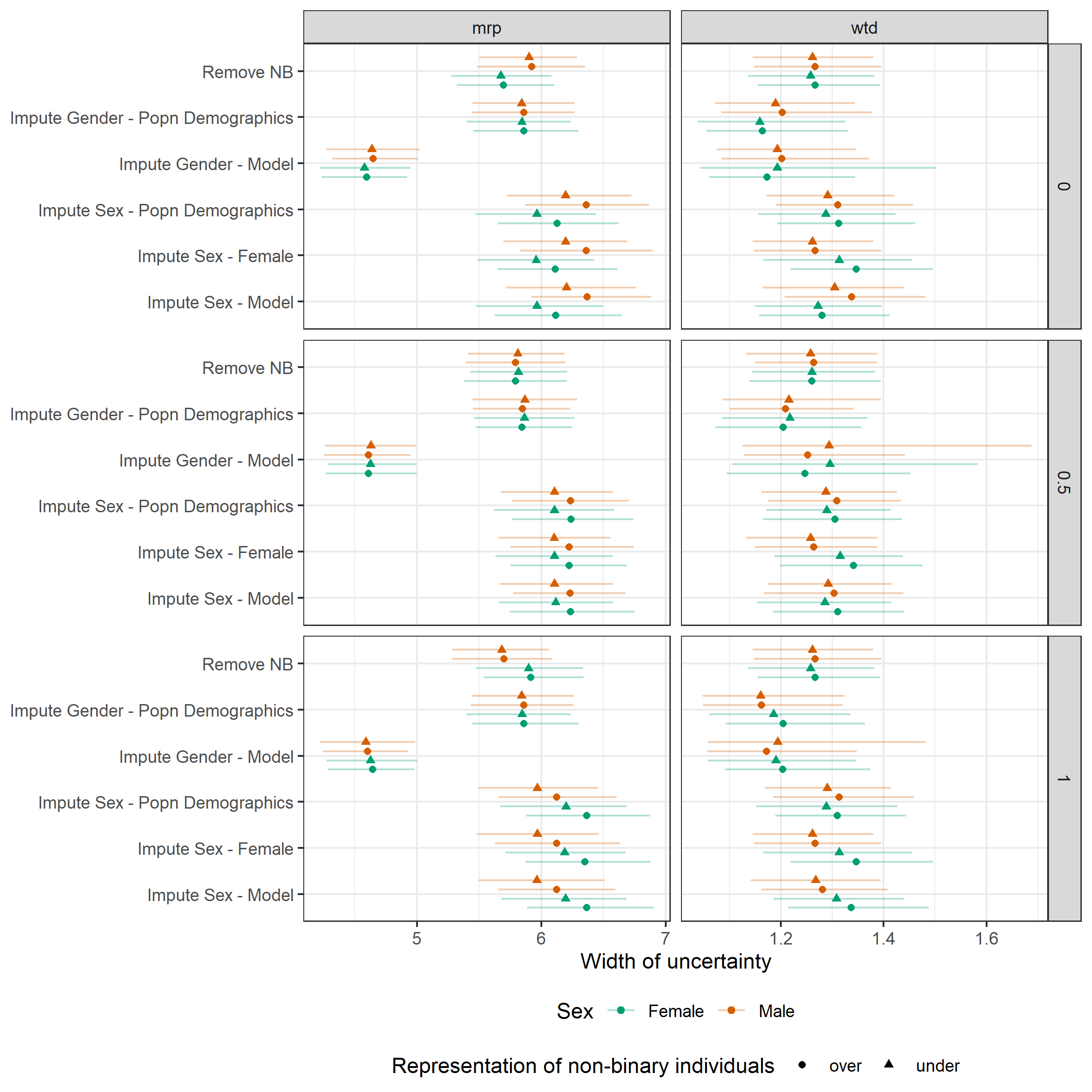}
    \caption{\em Width of uncertainty intervals of estimates of different sex demographics (M/F) when compared to population truth, where non-binary genders have different expected values on the outcome. The three row facets represent the different proportions of non-binary individuals who respond Male in response to a sex question. The two columns represent the analysis method (MRP and poststratification weights). Points represent the mean bias, with lines representing the 95\% quantiles across the 500 simulated data sets. Colour represents the sex demographic being predicted, while shape represents whether non-binary individuals were over or under represented in the data. These findings were not reported in the main text, but they largely reflect the pattern in uncertainty intervals as shown in the all different response categories (Figure 4).}
\end{figure}

\subsubsection{Gender demographic estimation}

\begin{figure}[H]
    \centering
    \includegraphics[width=\textwidth]{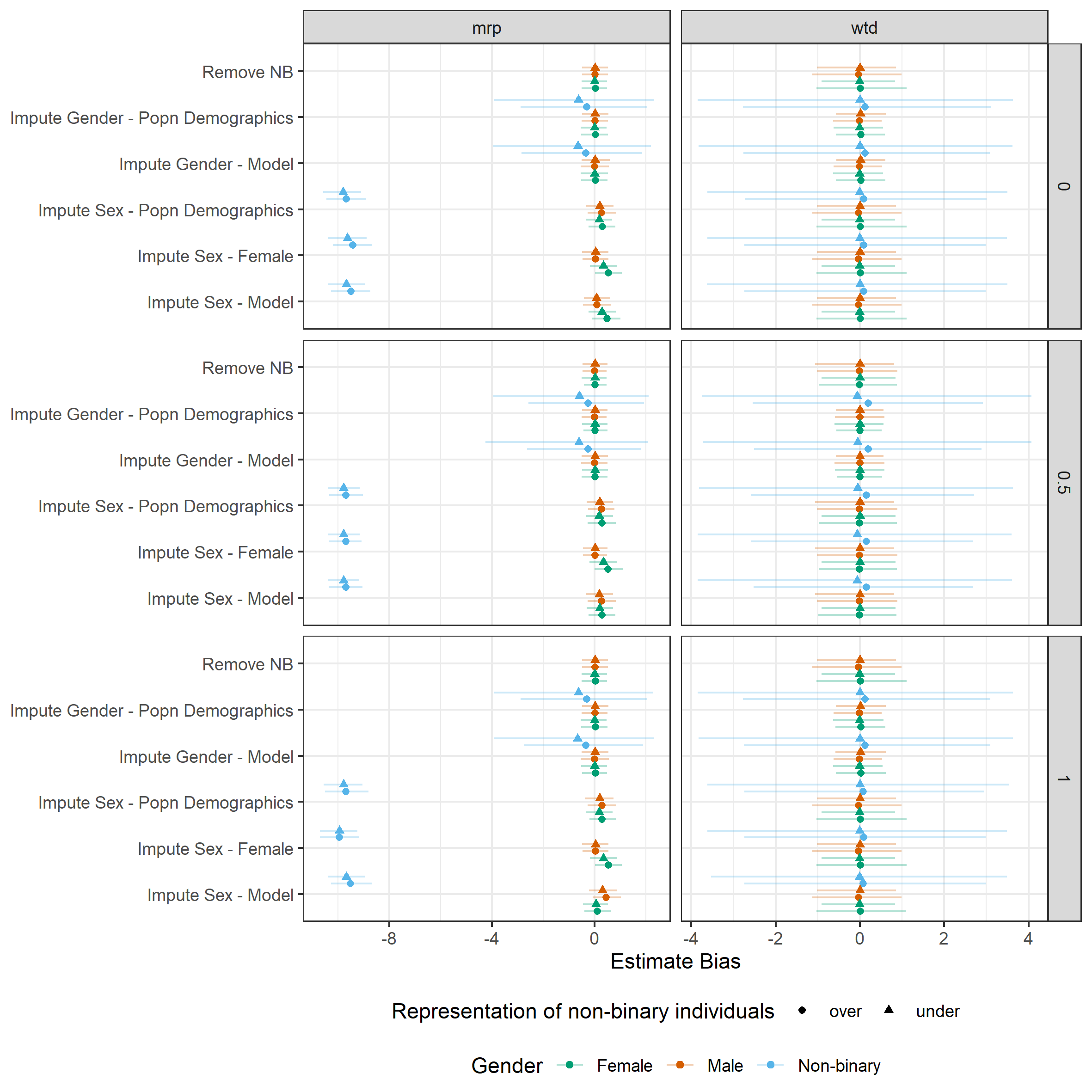}
    \caption{\em Bias of estimates of different gender demographics (M/F/NB) when compared to population truth, where non-binary genders have different expected values on the outcome. The three row facets represent the different proportions of non-binary individuals who respond Male in response to a sex question. The two columns represent the analysis method (MRP and poststratification weights). Points represent the mean bias, with lines representing the 95\% quantiles across the 500 simulated data sets. Colour represents the gender demographic being predicted, while shape represents whether non-binary individuals were over or under represented in the data. These findings were not reported in the main text. They reflect the differences in MRP approach, with imputing the sex methods with MRP relying essentially on the goodness of NB response to the sex question predictions (a difficult thing to model). The weighted estimates are all predominately unbiased, because the sex/gender variables do not greatly interact in predicting the outcome. }
\end{figure}

\begin{figure}[H]
    \centering
    \includegraphics[width=\textwidth]{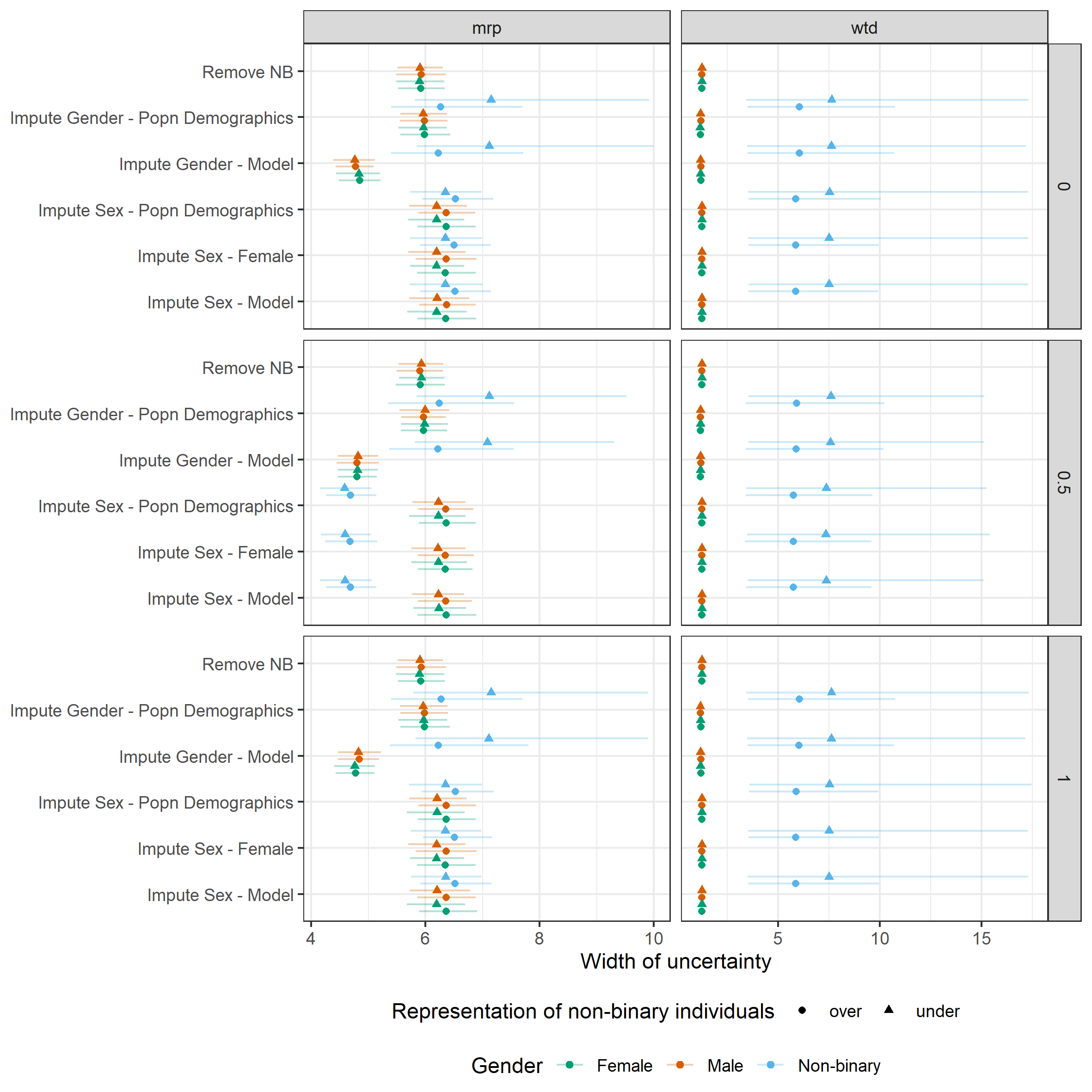}
    \caption{\em Width of uncertainty intervals of estimates of different gender demographics (M/F/NB) when compared to population truth, where non-binary genders have different expected values on the outcome. The three row facets represent the different proportions of non-binary individuals who respond Male in response to a sex question. The two columns represent the analysis method (MRP and poststratification weights). Points represent the mean bias, with lines representing the 95\% quantiles across the 500 simulated data sets. Colour represents the gender demographic being predicted, while shape represents whether non-binary individuals were over or under represented in the data.  These findings were not reported in the main text. They also largely represent the difference between MRP and weighted approaches, with weighted uncertainty much wider for the very small non-binary group. In MRP the differences are noisier, which could potentially be a reflection of the model mis-specification observed in Figure 11.}
\end{figure}

\subsection{Non-binary individuals simulated as responding similarly to female respondents, but differently to male respondents.}

\subsubsection{Population estimation}

\begin{figure}[H]
    \centering
    \includegraphics[width=\textwidth]{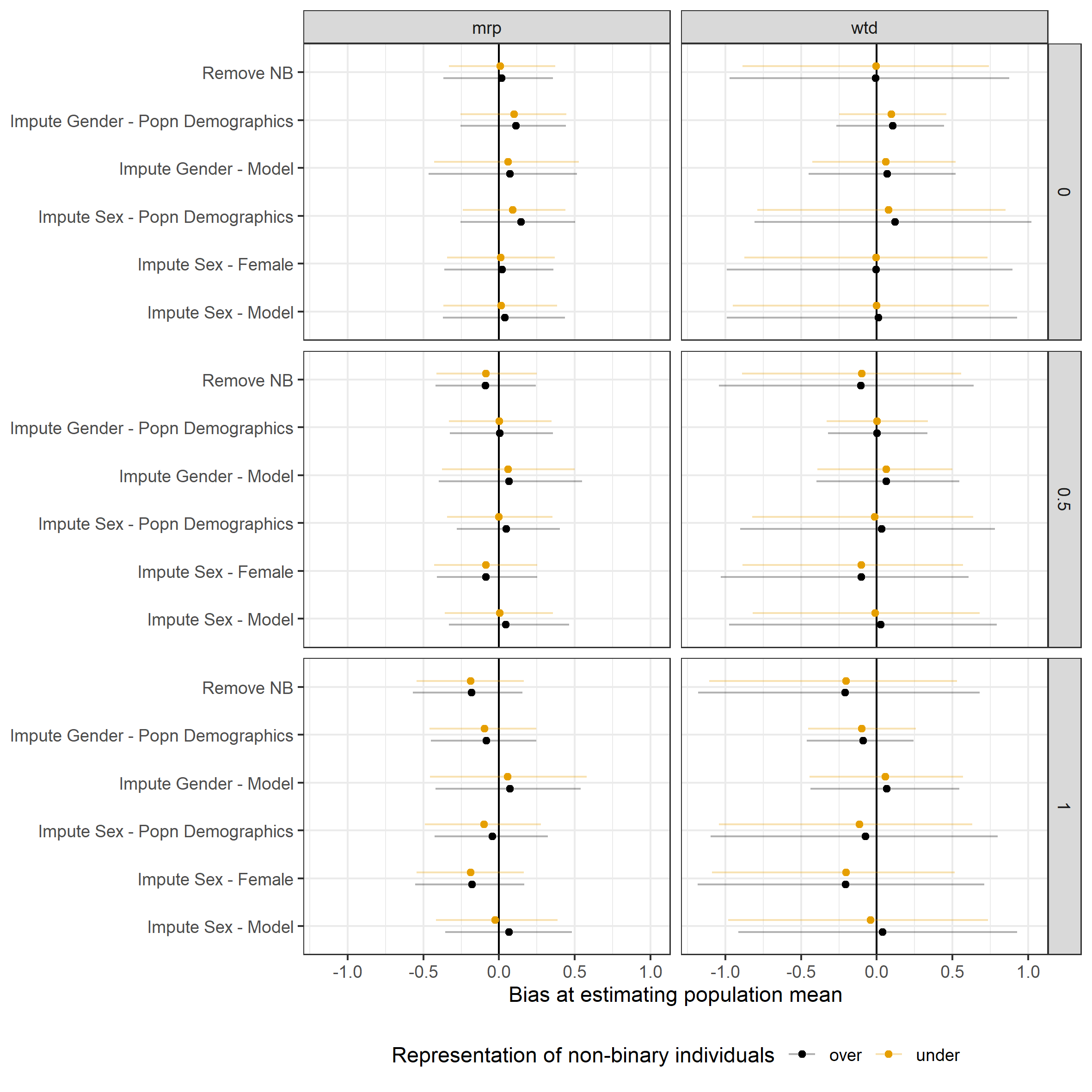}
    \caption{\em Bias of the MRP and weighted estimates from the sample, where non-binary genders are simulated to respond similarly as those of female gender. The three row facets represent the different proportions of non-binary individuals who respond Mal in response to a sex question. The two columns represent the analysis method (MRP and poststratification weights). Points represent the mean bias, with lines representing the 95\% quantiles across the 500 simulated data sets. Colour represents whether non-binary individuals were over or under represented in the data. This condition was not reported in the main text, but demonstrates overall unbiasedness of most methods in this scenario. }
\end{figure}

\begin{figure}[H]
    \centering
    \includegraphics[width=\textwidth]{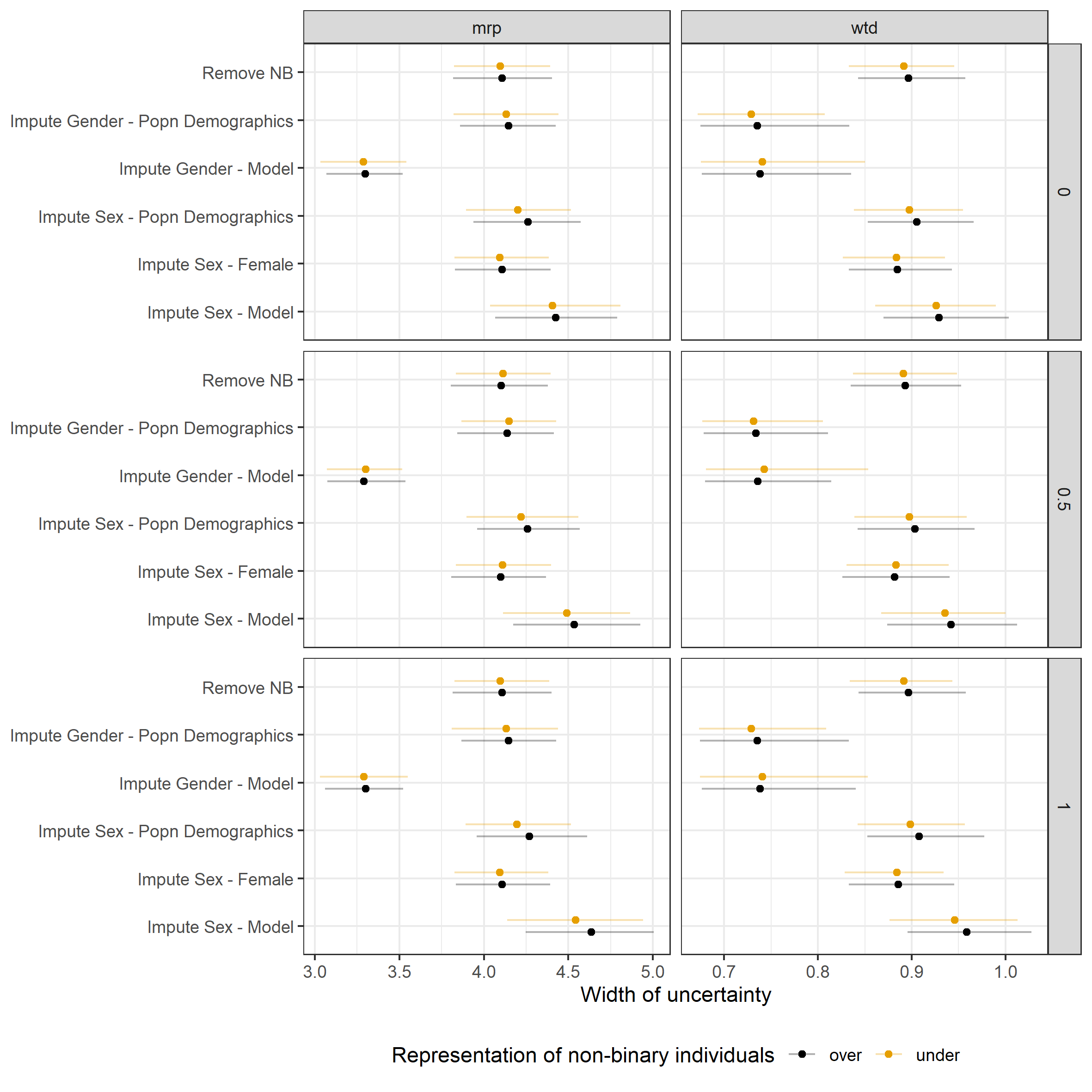}
    \caption{\em Width of uncertainty interval of the MRP and weighted estimates from the sample, where non-binary genders are simulated to respond similarly as those of female gender. The three row facets represent the different proportions of non-binary individuals who respond Male in response to a sex question. The two columns represent the analysis method (MRP and poststratification weights). Points represent the mean bias, with lines representing the 95\% quantiles across the 500 simulated data sets. Colour represents whether non-binary individuals were over or under represented in the data. This condition was not reported in the main text, but represents similar findings.}
\end{figure}

\subsubsection{Sex demographic estimation}

\begin{figure}[H]
    \centering
    \includegraphics[width=\textwidth]{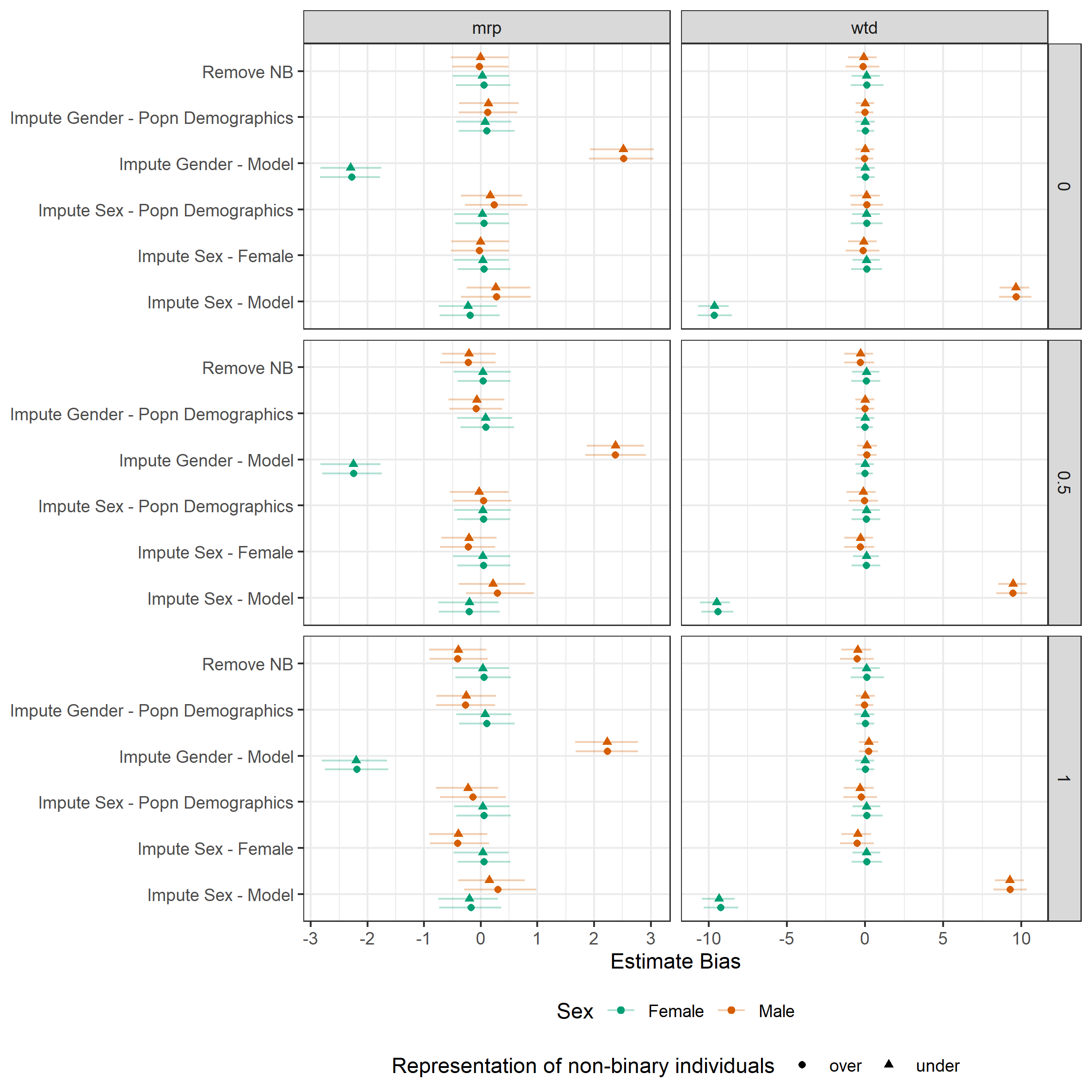}
    \caption{\em Bias of estimates of different sex demographics (M/F) when compared to population truth, where non-binary genders are simulated to respond similarly as those of female gender. The three row facets represent the different proportions of non-binary individuals who respond Male in response to a sex question. The two columns represent the analysis method (MRP and poststratification weights). Points represent the mean bias, with lines representing the 95\% quantiles across the 500 simulated data sets. Colour represents the sex demographic being predicted, while shape represents whether non-binary individuals were over or under represented in the data. These findings were not reported in the main text, but suggest a similar pattern. }
\end{figure}

\begin{figure}[H]
    \centering
    \includegraphics[width=\textwidth]{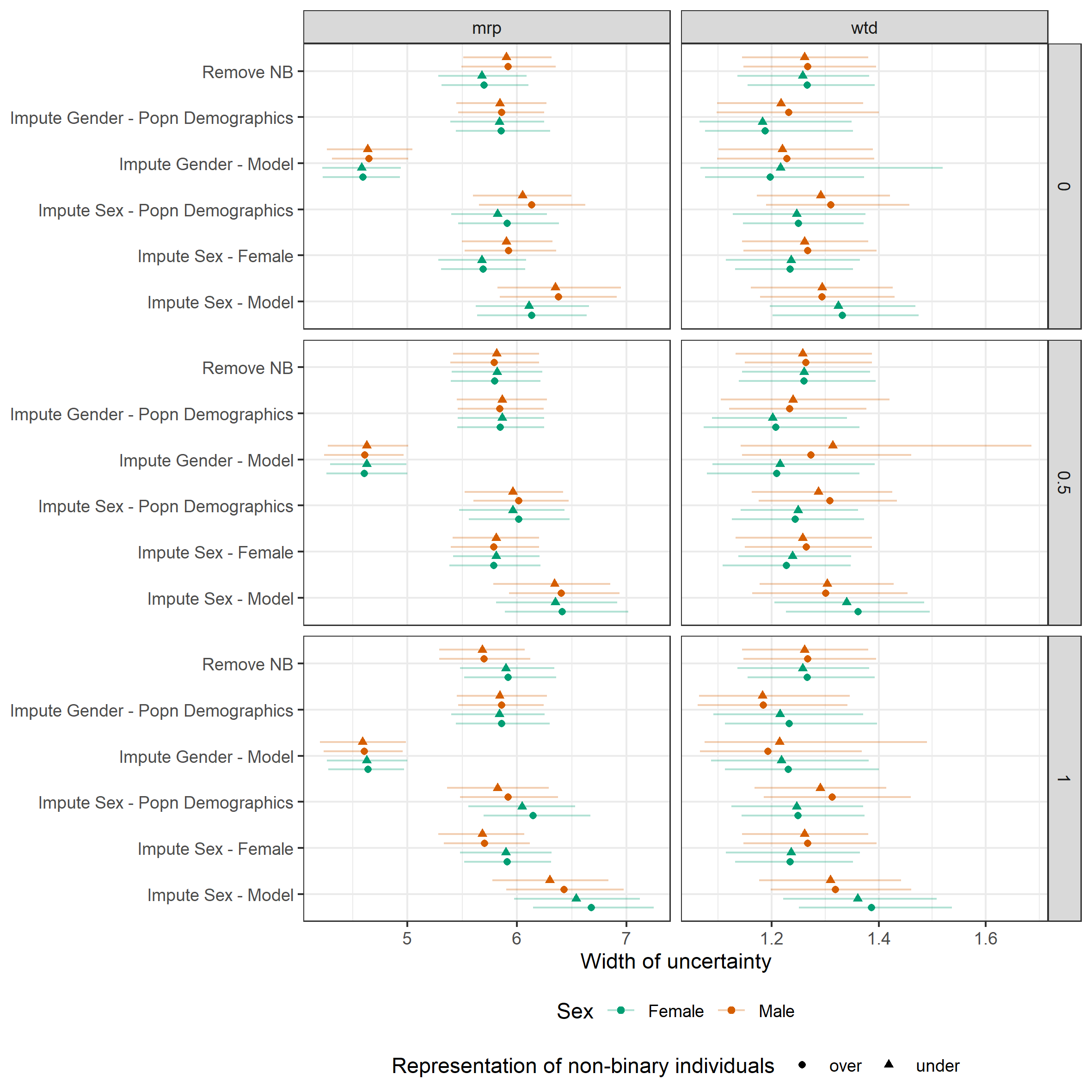}
    \caption{\em Width of uncertainty intervals of estimates of different sex demographics (M/F) when compared to population truth, where non-binary genders are simulated to respond similarly  as those of female gender. The three row facets represent the different proportions of non-binary individuals who respond Male in response to a sex question. The two columns represent the analysis method (MRP and poststratification weights). Points represent the mean bias, with lines representing the 95\% quantiles across the 500 simulated data sets. Colour represents the sex demographic being predicted, while shape represents whether non-binary individuals were over or under represented in the data. These findings were not reported in the main text, but reflect a similar pattern.}
\end{figure}

\subsubsection{Gender demographic estimation}

\begin{figure}[H]
    \centering
    \includegraphics[width=\textwidth]{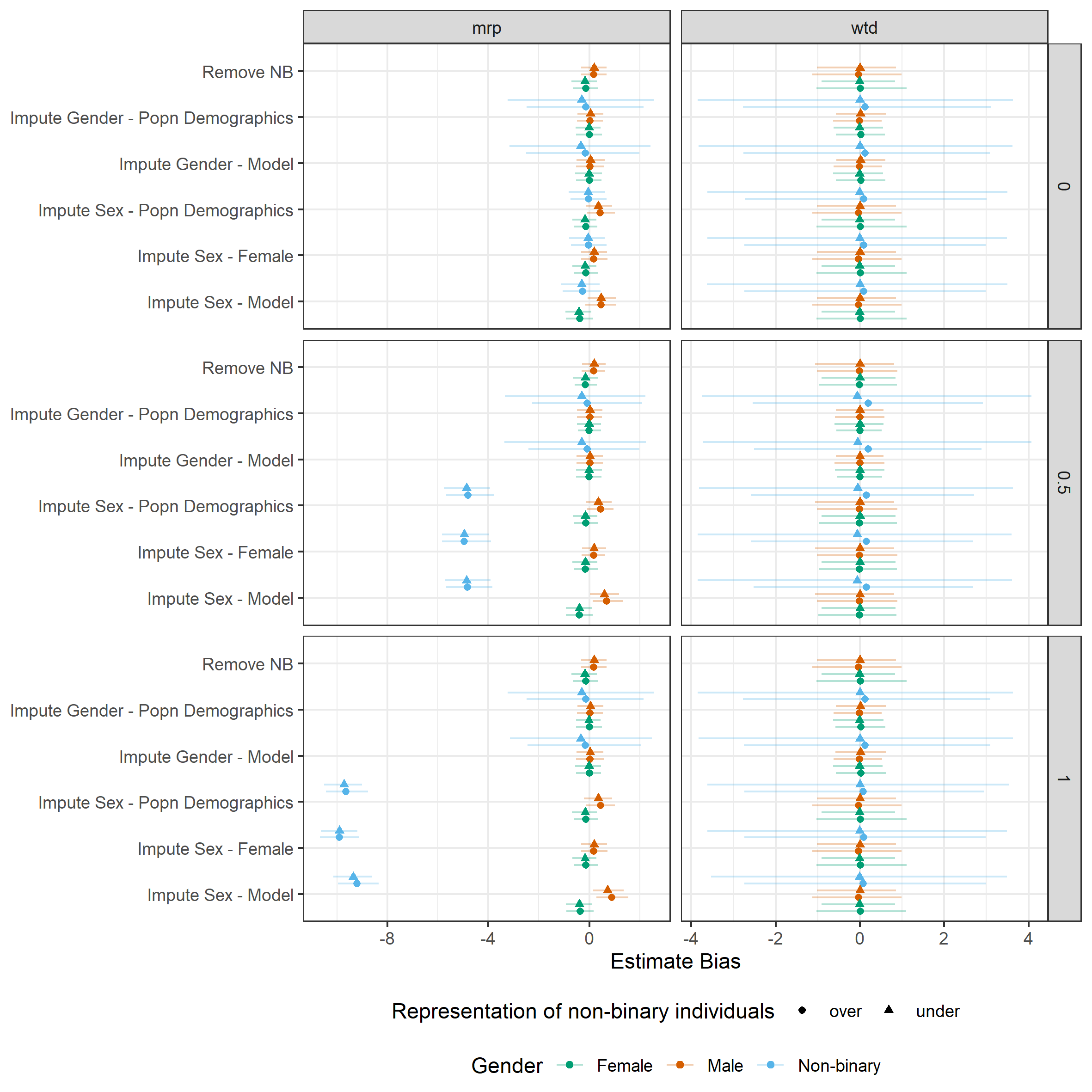}
    \caption{\em Bias of estimates of different gender demographics (M/F/NB) when compared to population truth, where non-binary genders are simulated to respond similarly as those of female gender. The three row facets represent the different proportions of non-binary individuals who respond Male in response to a sex question. The two columns represent the analysis method (MRP and poststratification weights). Points represent the mean bias, with lines representing the 95\% quantiles across the 500 simulated data sets. Colour represents the gender demographic being predicted, while shape represents whether non-binary individuals were over or under represented in the data. These findings were not reported in the main text. Similar to the previous scenario, imputing sex results in bias in the non-binary estimates, but only in the MRP scenarios.}
\end{figure}

\begin{figure}[H]
    \centering
    \includegraphics[width=\textwidth]{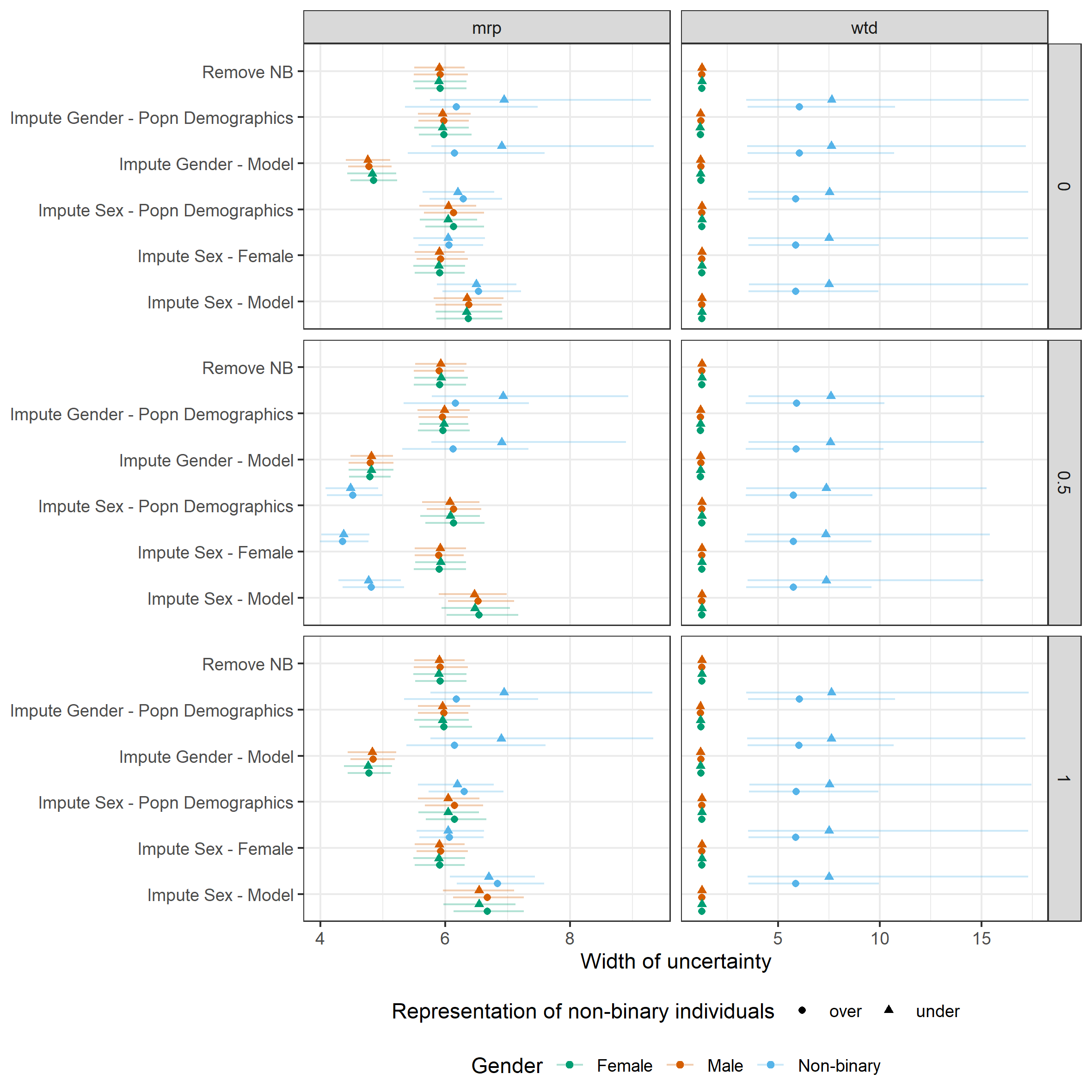}
    \caption{\em Width of uncertainty intervals of estimates of different gender demographics (M/F/NB) when compared to population truth, where non-binary genders are simulated to respond similarly as those of female gender. The three row facets represent the different proportions of non-binary individuals who respond Male in response to a sex question. The two columns represent the analysis method (MRP and poststratification weights). Points represent the mean bias, with lines representing the 95\% quantiles across the 500 simulated data sets. Colour represents the gender demographic being predicted, while shape represents whether non-binary individuals were over or under represented in the data.  These findings were not reported in the main text, but are similar to the findings from Section 2.}
\end{figure}

\subsection{No differences between the genders}

\subsubsection{Population estimation}

\begin{figure}[H]
    \centering
    \includegraphics[width=\textwidth]{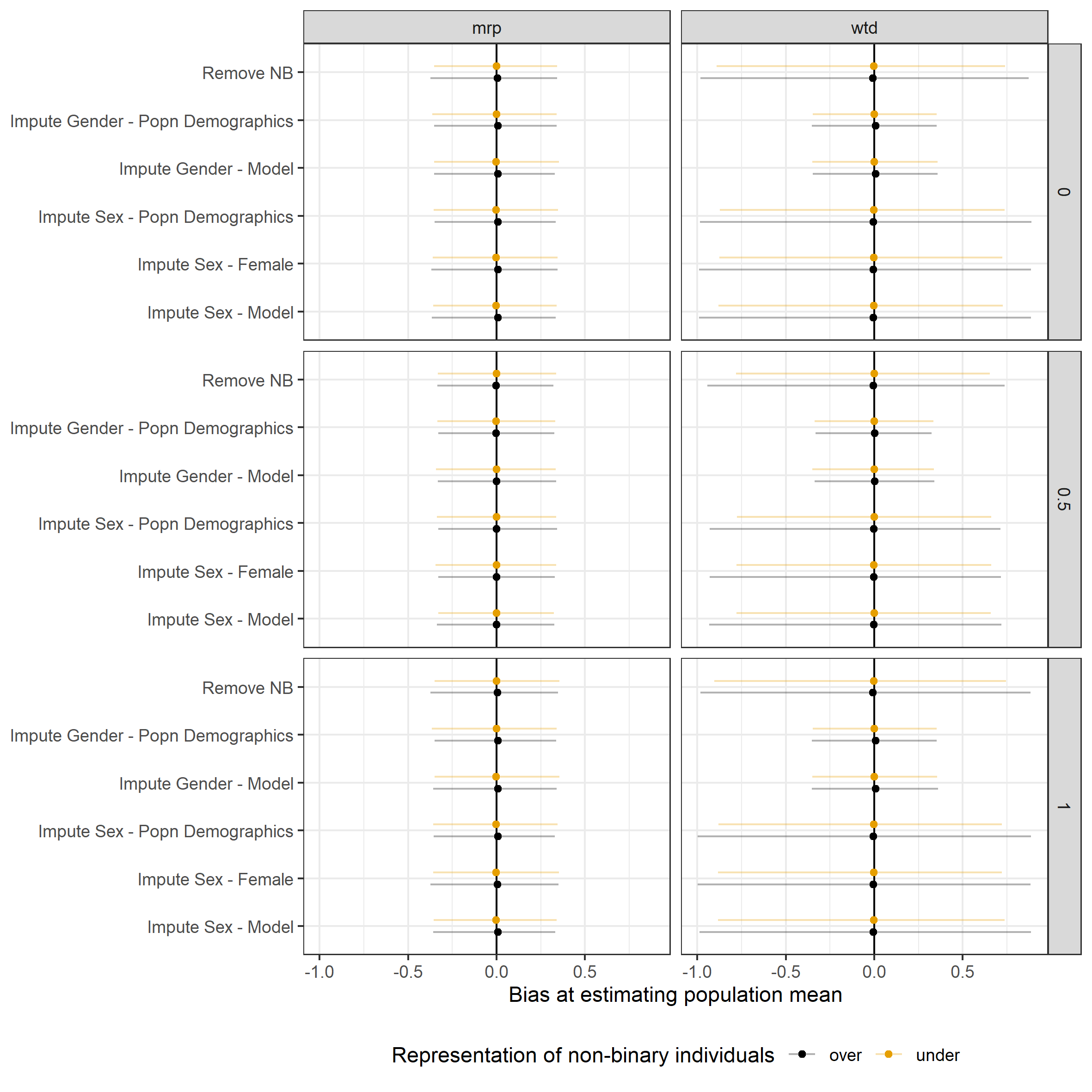}
    \caption{\em Bias of the MRP and weighted estimates from the sample, where gender is not related to the outcome. The three row facets represent the different proportions of non-binary individuals who respond Male in response to a sex question. The two columns represent the analysis method (MRP and poststratification weights). Points represent the mean bias, with lines representing the 95\% quantiles across the 500 simulated data sets. Colour represents whether non-binary individuals were over or under represented in the data. This condition was not reported in the main text, and demonstrates little difference between methods in terms of bias, with the only difference being a decrease in variance in the gender imputation based methods with weights.}
\end{figure}

\begin{figure}[H]
    \centering
    \includegraphics[width=\textwidth]{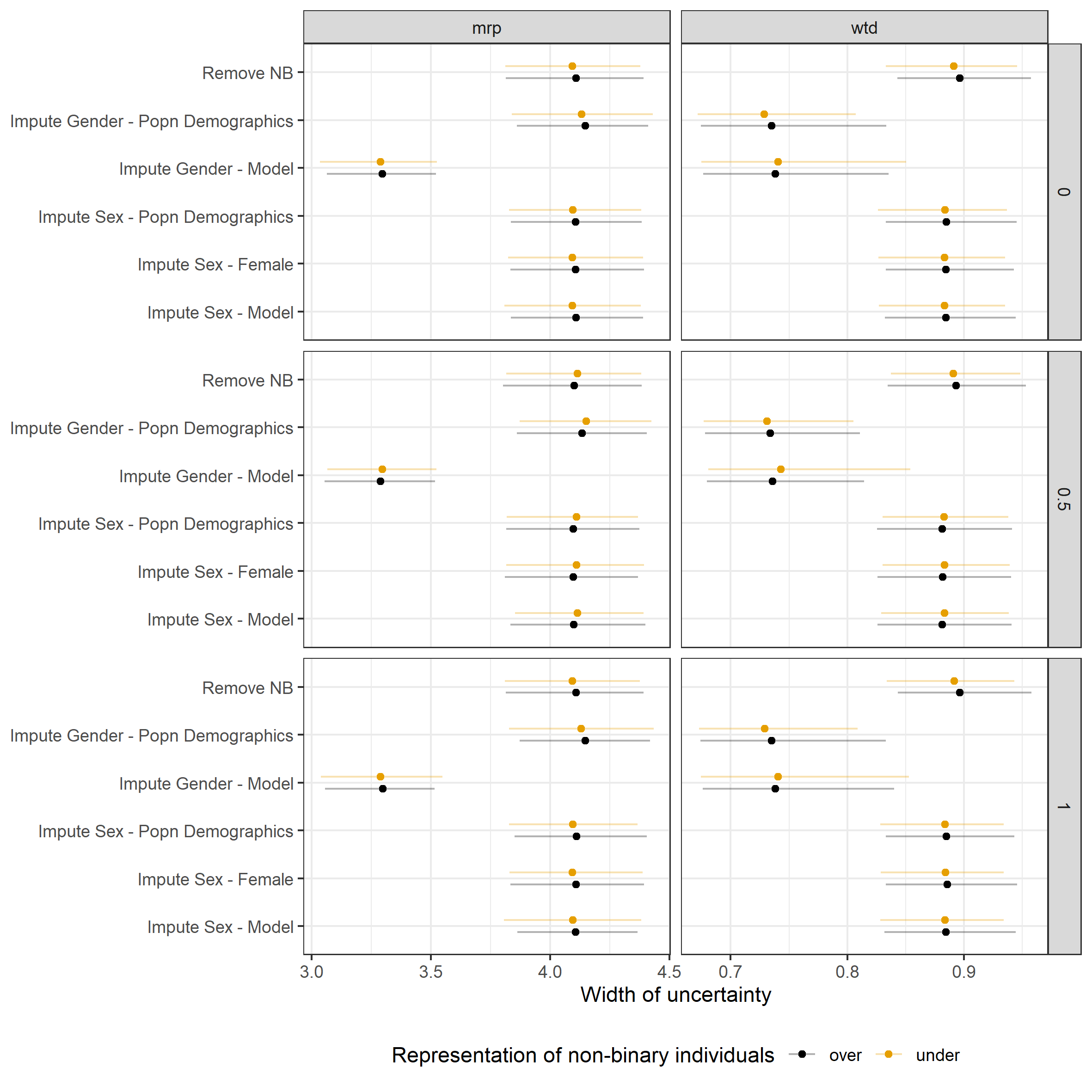}
    \caption{\em Width of uncertainty interval of the MRP and weighted estimates from the sample, where gender is not related to the outcome. The three row facets represent the different proportions of non-binary individuals who respond Male in response to a sex question. The two columns represent the analysis method (MRP and poststratification weights). Points represent the mean bias, with lines representing the 95\% quantiles across the 500 simulated data sets. Colour represents whether non-binary individuals were over or under represented in the data. This condition was not reported in the main text, but represents similar estimates of uncertainty, indicating that this might be a feature of the method rather than the data. }
\end{figure}

\subsubsection{Sex demographic estimation}

\begin{figure}[H]
    \centering
    \includegraphics[width=\textwidth]{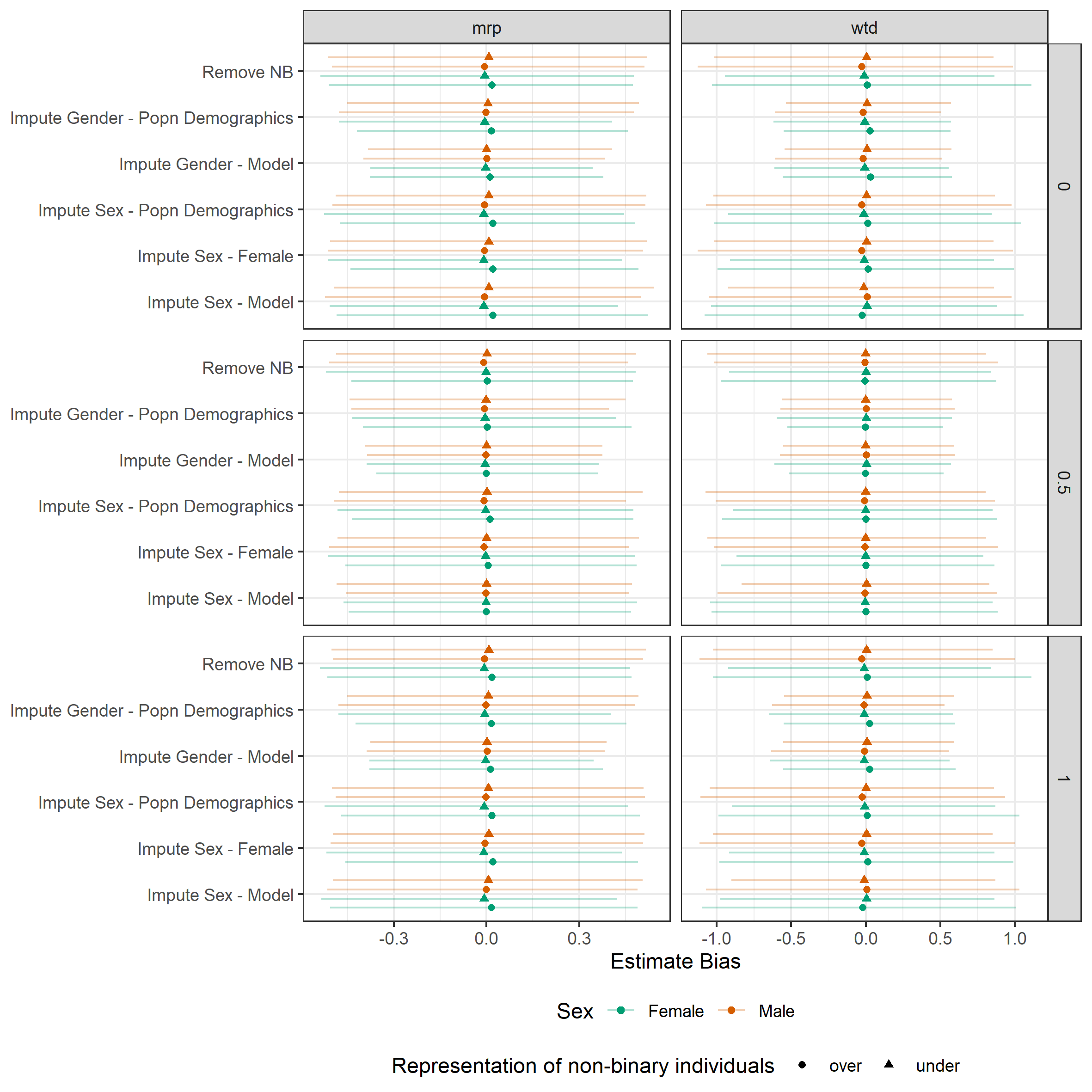}
    \caption{\em Bias of estimates of different sex demographics (M/F) when compared to population truth, where gender is not related to the outcome. The three row facets represent the different proportions of non-binary individuals who respond Male in response to a sex question. The two columns represent the analysis method (MRP and poststratification weights). Points represent the mean bias, with lines representing the 95\% quantiles across the 500 simulated data sets. Colour represents the sex demographic being predicted, while shape represents whether non-binary individuals were over or under represented in the data. These findings were not reported in the main text, but suggest little to no difference between methods, except the same reduction in variance observed in Figure 19. }
\end{figure}

\begin{figure}[H]
    \centering
    \includegraphics[width=\textwidth]{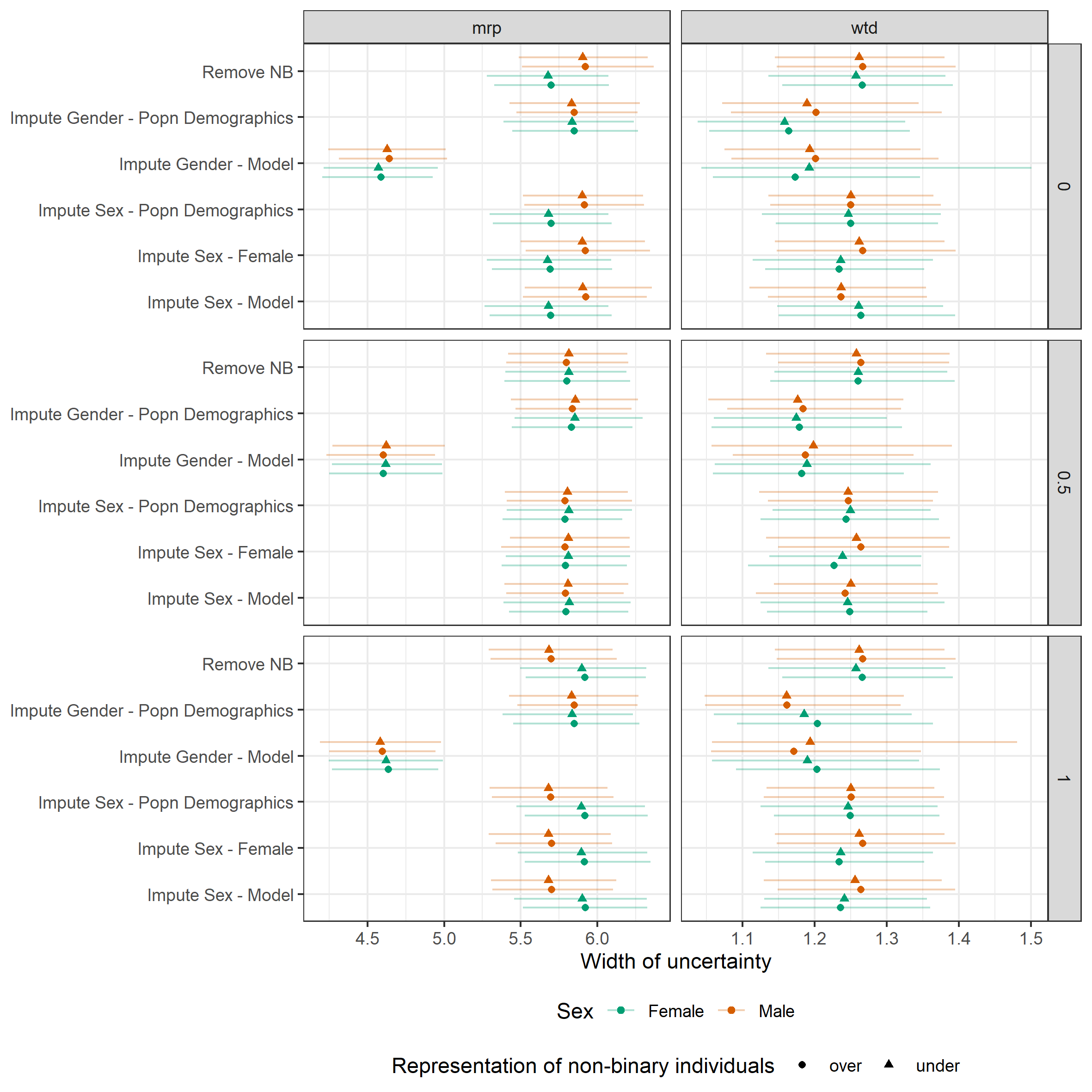}
    \caption{\em Width of uncertainty intervals of estimates of different sex demographics (M/F) when compared to population truth, where gender is not related to the outcome. The three row facets represent the different proportions of non-binary individuals who respond Male in response to a sex question. The two columns represent the analysis method (MRP and poststratification weights). Points represent the mean bias, with lines representing the 95\% quantiles across the 500 simulated data sets. Colour represents the sex demographic being predicted, while shape represents whether non-binary individuals were over or under represented in the data. These findings were not reported in the main text, but relatively similar uncertainty in all methods and conditions, except the impute gender with a model method, which seems more confident (remember imputation uncertainty was not included in any method).}
\end{figure}

\subsubsection{Gender demographic estimation}

\begin{figure}[H]
    \centering
    \includegraphics[width=\textwidth]{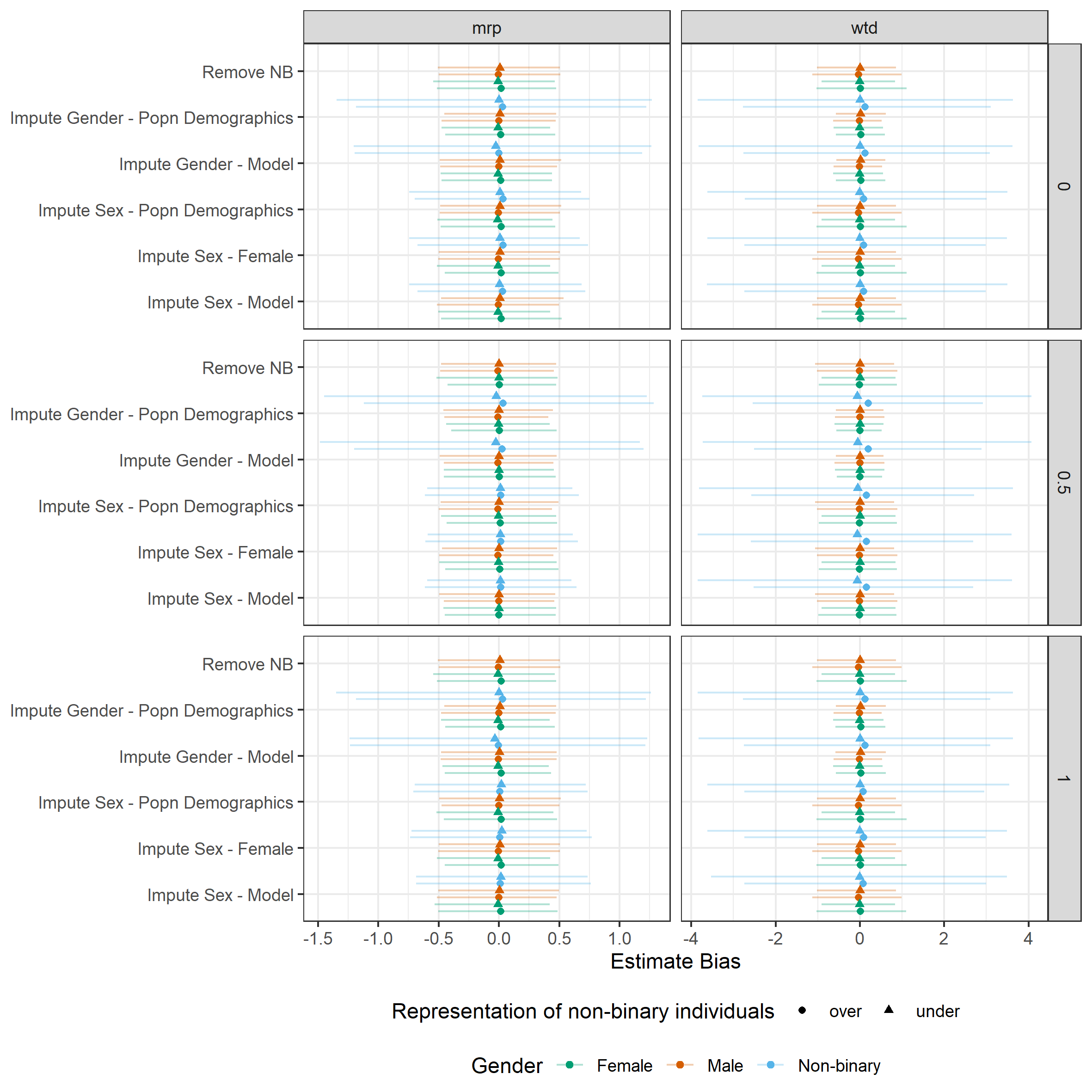}
    \caption{\em Bias of estimates of different gender demographics (M/F/NB) when compared to population truth, where gender is not related to the outcome. The three row facets represent the different proportions of non-binary individuals who respond Male in response to a sex question. The two columns represent the analysis method (MRP and poststratification weights). Points represent the mean bias, with lines representing the 95\% quantiles across the 500 simulated data sets. Colour represents the gender demographic being predicted, while shape represents whether non-binary individuals were over or under represented in the data. These findings were not reported in the main text. There a little differences between methods in terms of bias, with the non-binary estimates having greater variance in bias across simulation iterations. }
\end{figure}

\begin{figure}[H]
    \centering
    \includegraphics[width=\textwidth]{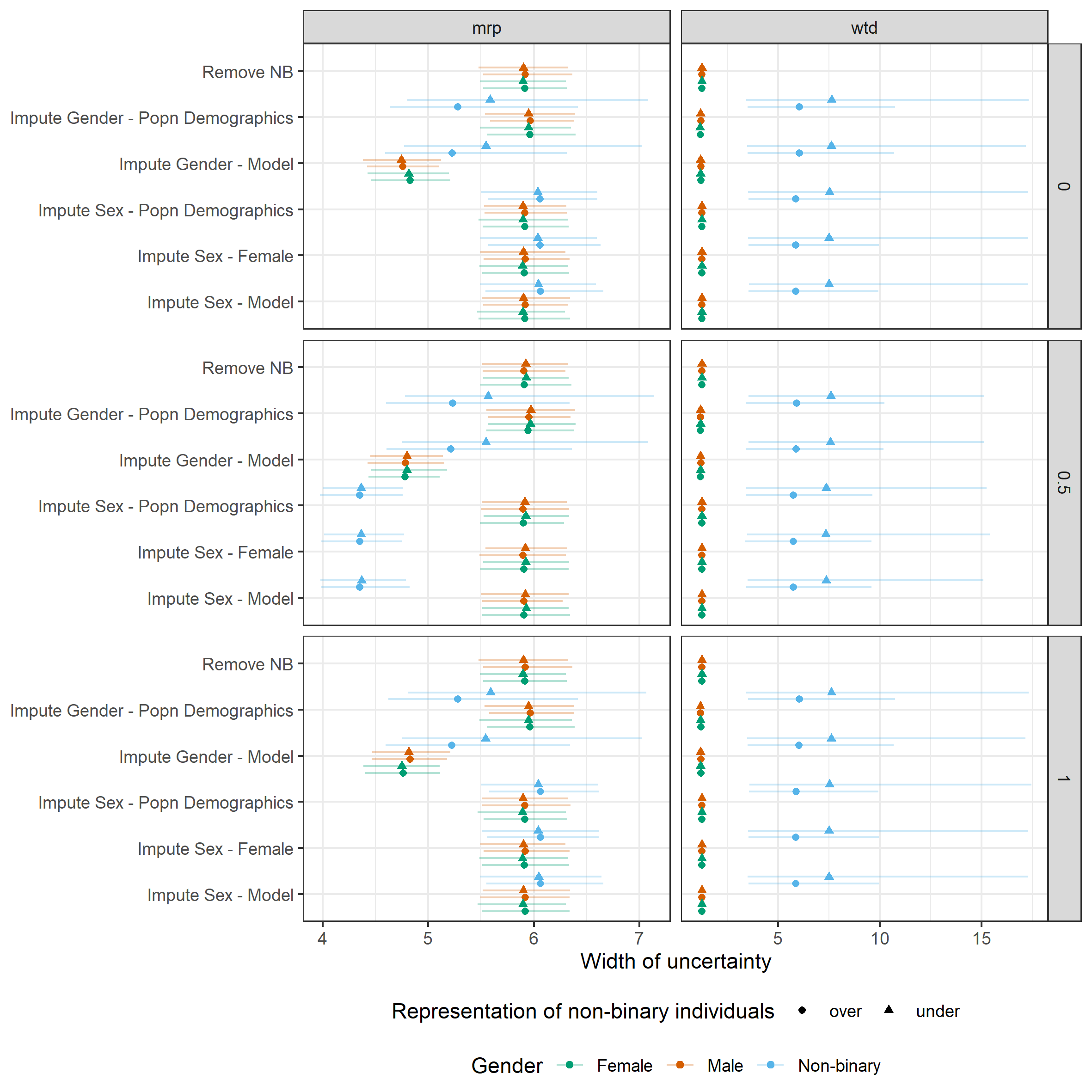}
    \caption{\em Width of uncertainty intervals of estimates of different gender demographics (M/F/NB) when compared to population truth, where gender is not related to the outcome. The three row facets represent the different proportions of non-binary individuals who respond Male in response to a sex question. The two columns represent the analysis method (MRP and poststratification weights). Points represent the mean bias, with lines representing the 95\% quantiles across the 500 simulated data sets. Colour represents the gender demographic being predicted, while shape represents whether non-binary individuals were over or under represented in the data.  These findings were not reported in the main text, but reflect differences between MRP and weight-based methods. In weight-based methods, the uncertainty for non-binary individuals was relatively similar across methods, whereas for MRP-based estimates the uncertainty reflects the model used. Models with sex as a predictor were more confident in non-binary estimates than models with gender as a predictor.}
\end{figure}

\end{document}